\documentclass[11pt,a4paper]{article}
\usepackage[T1]{fontenc}
\usepackage{amssymb}
\usepackage{mathptmx}
\usepackage[pdftex]{graphicx}
\usepackage[english]{babel}
\usepackage[pdftex,linkcolor=black,pdfborder={0 0 0}]{hyperref}
\usepackage{calc} 
\usepackage{enumitem} 
\usepackage{setspace}
\usepackage[a4paper, lmargin=0.12\paperwidth, rmargin=0.12\paperwidth, tmargin=0.1111\paperheight, bmargin=0.1111\paperheight]{geometry} 
\usepackage{wallpaper}
\usepackage{mdframed}
\usepackage{tikz}
\usepackage{comment}
\usepackage[affil-it]{authblk}
\usepackage{siunitx}
\usepackage{amsmath}
\usepackage{xr-hyper} 
\usepackage{comment}

\usepackage[all]{nowidow} 
\usepackage[protrusion=true,expansion=true]{microtype} 
\usepackage{csquotes}

\usepackage[backend=biber,style=ieee, autocite=superscript]{biblatex} 
\author[1,*]{Nicola Melchioni}
\author[1]{Andrea Mancini}
\author[1,*]{Antonio Ambrosio}

\affil[1]{Centre for Nano Science and Technology, Fondazione Istituto Italiano di Tecnologia, Via Rubattino 81, Milano 20134, Italy}
\affil[*]{Corresponding authors: nicola.melchioni@iit.it, antonio.ambrosio@iit.it}

\makeatletter
\title{Anisotropic electron gas in a hyperbolic van der Waals material}

\hypersetup{ 	
pdfsubject = {},
pdftitle = {Anisotropic electron gas in a hyperbolic van der Waals material},
pdfauthor = {}
}

\newcommand*{\addFileDependency}[1]{
\typeout{(#1)}
\@addtofilelist{#1}
\IfFileExists{#1}{}{\typeout{No file #1.}}
}\makeatother

\begin{document}

\maketitle
\doublespacing
\newpage
\section*{Abstract}
Electron gases in low-dimensional materials exhibit unconventional transport and optical phenomena due to reduced phase space, enhanced interactions, and strong sensitivity to lattice symmetry. While commonly realized in quantum-confined systems and engineered heterostructures, such states are rare in naturally occurring materials. Hyperbolic materials provide a compelling alternative, as extreme lattice anisotropy can host unconventional electronic states and novel electron–phonon interactions. Here, we investigate the angle-resolved polarized Raman (ARPR) response of MoOCl\textsubscript{2}, the first naturally occurring hyperbolic material whose hyperbolicity originates from a highly anisotropic electron gas. We observe pronounced polarization-dependent Fano line shapes, revealing coherent coupling between phonons and an anisotropic electronic continuum. We characterize the directional response of this continuum, incorporating it into effective Raman tensors that quantitatively reproduce the ARPR measurements and capture the distinct Raman fingerprint of MoOCl\textsubscript{2}. Excitation-energy and thickness-dependent ARPR measurements further demonstrate a tunable quasi-1D electronic continuum with weak interlayer coupling, establishing MoOCl\textsubscript{2} as a model system for Raman studies of electron–phonon coupling in hyperbolic materials.

\newpage

Hyperbolic materials, characterized by dielectric tensor components of opposite sign, have emerged as powerful platforms for manipulating light–matter interactions at the nanoscale \cite{poddubny2013hyperbolic, takayama2019optics}. Early implementations relied on artificial metamaterials composed of alternating metallic and dielectric elements \cite{shekhar2014hyperbolic}. Although these systems enabled landmark demonstrations such as negative refraction \cite{high2015visible} and hyperlensing \cite{lee2017realization}, their performance is ultimately limited by fabrication constraints and significant optical losses associated with metals \cite{poddubny2013hyperbolic}. In addition, the hyperbolic behavior of artificially engineered metamaterials arises from an effective macroscopic response, rather than from intrinsically anisotropic light–matter interactions at the atomic scale.

By contrast, in natural hyperbolic crystals, the optical response emerges directly from intrinsic electronic or lattice dynamics. In this regard, the broad family of van der Waals (vdW) materials provides different examples of hyperbolicity \cite{hu2020phonon, wang2024planar}. For instance, in MoO\textsubscript{3} \cite{ma2018plane, zheng2019mid} and hexagonal boron nitride (hBN) \cite{dai2014tunable, ambrosio2018selective}, hyperbolicity originates from strongly anisotropic optical phonons, giving rise to reststrahlen bands that support hyperbolic phonon polaritons. In these materials, where the hyperbolic response is typically confined to the infrared due to the characteristic phonon energies, the vibrational origin of hyperbolicity provides direct access to the underlying anisotropic lattice dynamics, which has been probed using different near- and far-field techniques \cite{galiffi2024extreme}, including Raman spectroscopy \cite{bergeron2023probing}.

More recently, natural hyperbolicity has been extended to the visible spectral range by MoOCl\textsubscript{2} \cite{zhao2020highly, venturi2024visible, ruta2024good}. Unlike previous materials, where the optical response is mainly dictated by lattice vibrations, in MoOCl\textsubscript{2} the hyperbolic response arises from the polarizability of an electron gas with strongly anisotropic effective masses, which results in markedly different plasma frequencies along the crystallographic axes \cite{zhao2020highly, melchioni2025giant, ghosh2025direct}. This type of hyperbolicity offers a unique platform in which the interplay between electronic anisotropy and electron–phonon interactions can be explored. Such interactions can provide unprecedented insight into the microscopic mechanisms governing light–matter coupling in strongly anisotropic systems.

In this work, we investigate MoOCl\textsubscript{2} using angle-resolved polarized Raman spectroscopy (ARPR), revealing coupling between phonons and a strongly anisotropic electronic continuum. We find that the Raman peaks selectively interact with the continuum, giving rise to polarization-dependent Fano lineshapes. By extending the tensorial description of Raman scattering to account for the continuum response, we extract the full angle-dependent lineshape of the peaks. We demonstrate that by varying the excitation energy, the strength of the electron–phonon coupling can be actively tuned, providing a direct handle on the interplay between lattice vibrations and the anisotropic electronic continuum. Finally, by recording ARPR spectra on flakes of varying thickness, we show that the coupling strength evolves with thickness, providing a clear fingerprint of an electronic continuum with weak interlayer coupling. Taken together, our findings demonstrate that the electron gas in MoOCl\textsubscript{2} is intrinsically quasi-1D, predominantly confined along the Mo–O chains and weakly dispersive in the transverse directions. This establish MoOCl\textsubscript{2} as a unique platform where the combination of strong optical and electrical anisotropy, hyperbolic dispersion, and directionally selective electron–phonon interactions can be exploited for interesting applications in nanophotonics and nanochemistry, such as tailoring light–matter coupling, enhancing Raman responses, and engineering nanoscale energy transport.



\begin{figure}[!t]
    \centering
    \includegraphics[width=0.9\linewidth]{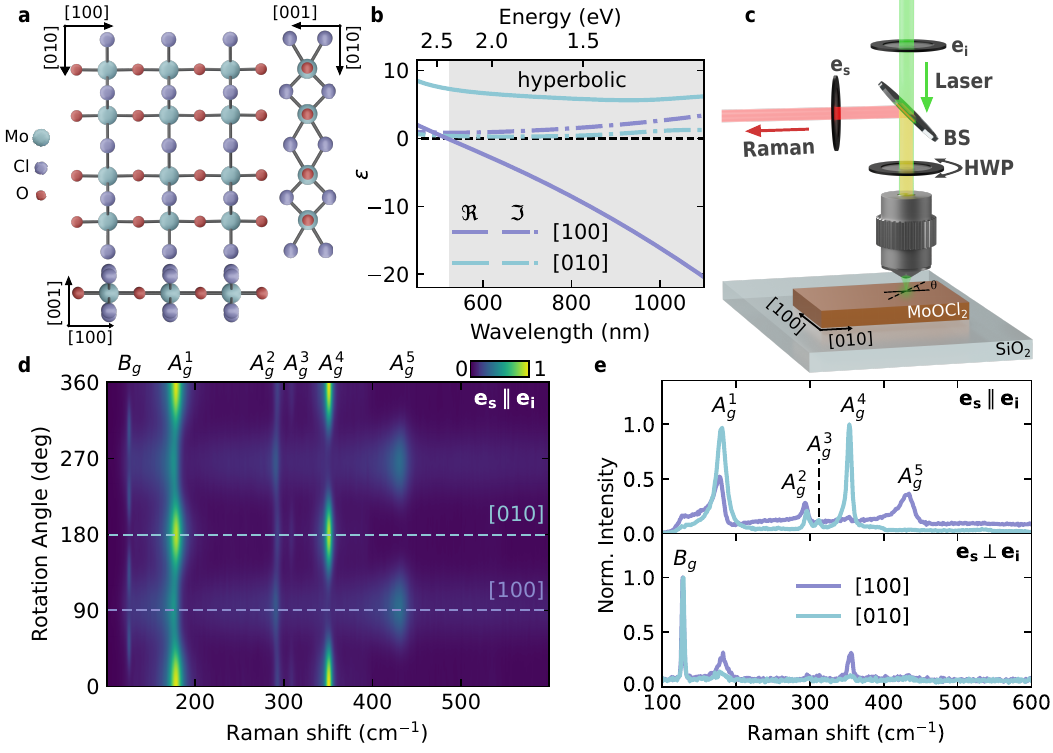}
    \caption{Angle-resolved polarized Raman measurements on MoOCl\textsubscript{2}. \textbf{a} Crystal structure of MoOCl\textsubscript{2} projected onto the three orthogonal planes. \textbf{b} Real part (continuous lines) and imaginary part (dash-dotted lines) of the in-plane dielectric tensor of MoOCl\textsubscript{2}. The region in which the material is hyperbolic is shaded in gray. The transition between hyperbolic and elliptic happens at \SI{\sim520}{\nm}. Model from \cite{melchioni2025giant}. \textbf{c} Schematics of the experimental ARPR setup. Light is initialized to $\mathbf{e_i}$, then rotated in-plane by the half wave plate (HWP). Scattered light is passed through the HWP and redirected by a beam-splitter (BS) to the analizer $\mathbf{e_s}$, effectively simulating a rotation of the sample under the microscope. \textbf{d} Normalized ARPR spectra of MoOCl\textsubscript{2} collected in parallel polarization with excitation energy \SI{532}{\nm}. Dashed lines indicate the orientation of the crystal axes. \textbf{e} Raman spectra collected with polarization along the [100] axis (purple lines) and the [010] axis (blue lines) with $e_s \parallel e_i$ (upper panel, extracted from (d)) and $e_s \perp e_i$ (lower panel) polarizations.}
    \label{fig1}
\end{figure}

The crystal structure of MoOCl\textsubscript{2}, shown in Fig. \ref{fig1}a, is characterized by a monoclinic lattice belonging to the C\textit{2/m} space group. The monolayer consists of metallic Mo–O chains aligned to the [100] direction, interconnected by Cl atoms \cite{zhao2020highly}. Along the [010] direction, the Cl atoms undergo dimerization driven by Peierls distortions \cite{zhang2021orbital, helmer2025computational}, leading to partial electron localization and semiconducting behavior. As a result of the distinct bonding character along the two in-plane directions, the electronic band structure of MoOCl\textsubscript{2} is strongly anisotropic, giving rise to strikingly different plasma frequencies along the principal axes \cite{zhao2020highly}. In particular, the plasma frequency along the [100] direction lies at approximately \SI{2.38}{\eV}. For photon energies below this threshold, MoOCl\textsubscript{2} exhibits an in-plane hyperbolic optical response, behaving as a metal ($\varepsilon_x < 0$) along [100] and as a dielectric ($\varepsilon_y > 0$) along [010] (shaded gray region in Fig. \ref{fig1}b). At higher photon energies, above the [100] plasma frequency, both in-plane dielectric tensor components become positive, and the material transitions to an elliptic optical response (white region of Fig. \ref{fig1}b).
Due to the crystal symmetry of MoOCl\textsubscript{2}, the Raman-active optical phonons can be classified into two irreducible representations, namely $A_g$ and $B_g$ modes \cite{loudon1964raman}. The $A_g$ modes correspond to vibrations that preserve the mirror symmetry of the lattice and are primarily associated with atomic displacements along the principal crystallographic axes, whereas the $B_g$ modes involve antisymmetric vibrations (see Supplementary Information S2 for a more detailed description and the corresponding Raman tensors).
To characterize the optical phonons of MoOCl\textsubscript{2}, we performed ARPR measurements on a \SI{305}{\nm}-thick flake using excitation wavelength of \SI{532}{\nm} (Fig. \ref{fig1}c). When measured with parallel polarizers, the spectra reveal six distinct Raman peaks (Fig. \ref{fig1}d, see Supplementary Information S1 for additional experimental details and S5 for measurements in cross polarization). The lowest-energy peak, centered at approximately \SI{126}{\cm^{-1}}, is only observed in parallel configuration when the incident polarization is not parallel to the principal crystallographic axes and vanishes when the polarization is aligned along either axis (see also Fig. \ref{fig1}e, where the peak is clearly visible when the experiment is repeated in cross polarization), consistent with the selection rules of a $B_g$ symmetry mode. In contrast, the remaining five peaks, centered respectively at \SI{180}{\cm^{-1}}, \SI{296}{\cm^{-1}}, \SI{311}{\cm^{-1}}, \SI{353}{\cm^{-1}} and \SI{434}{\cm^{-1}}, exhibit a different angular dependence, with intensity maxima aligned along the crystallographic axes, and are identified as modes of $A_g$ symmetry. Finally, the Raman signal away from the phonon peaks exhibits an enhanced baseline along [100] (purple dashed line in Fig. \ref{fig1}d), indicating the presence of a directional electronic continuum that contributes to the Raman response along that axis \cite{klein2005electronic}. The observed Raman peaks are consistent with those reported in recent studies \cite{ghosh2025direct, margaryan2025dielectric}, and with the experimental and theoretical results of Minnekhanov et al. \cite{minnekhanov2025hyperbolic}, which are based on a similar analysis. From now on, we will refer to the first peak as $B_g$, and label the remaining five peaks as $A_g^j$ with $j = 1,\dots,5$ (see Fig. \ref{fig1}e).

A striking feature of the Raman response of MoOCl\textsubscript{2} emerges when the incident and scattered light are aligned with the crystallographic axes. In the parallel polarization configuration, spectra acquired with the polarization oriented along the [010] direction (Fig.~\ref{fig1}e, blue line in the upper panel) exhibit nearly Lorentzian lineshapes, as typically observed for Raman scattering in dielectric media. By contrast, when the polarization is aligned along the [100] axis (purple line in the upper panel of Fig. \ref{fig1}e), the $A_g$ Raman peaks develop a pronounced asymmetry, indicative of a fundamentally different scattering mechanism. Such behavior has previously been observed in systems where a charge-carrier gas coherently interacts with lattice phonons \cite{magidson2002fano, zhang2022fano, tanwar2022fano}, giving rise to the characteristic Fano lineshape that results from the coherent coupling between a discrete resonance (the phonon) and a continuum \cite{bechstedt1975theory}. In contrast, the $B_g$ peak exhibits a markedly different behavior, retaining a Lorentzian lineshape regardless of the polarization orientation (lower panel of Fig. \ref{fig1}e). These observations confirm the presence of an anisotropic electronic continuum confined along the [100] axis of the crystal, whose interaction with light is revealed by the Raman process.

To characterize the asymmetry of the peaks and quantify the coupling strength of the phonon to the continuum, we fitted the peaks using Fano lineshapes, which can be expressed as
\begin{equation}
I(\omega) = I_0\frac{(q+\varepsilon)^2}{1+\varepsilon^2}, \qquad \mathrm{with}\ \varepsilon = \frac{\omega-\omega_0}{\gamma}
\end{equation}
where $\omega_0$ and $\gamma$ are the central frequency and broadening of the phonon resonance, respectively, and $q$ is the Fano parameter which quantifies the asymmetry of the peak. 
Using this model, both the $B_g$ and $A_g$ peaks can be fitted consistently, as shown in Fig. \ref{fig2}a for the $B_g$ and $A_g^1$ peaks (see Supplementary Information S3 for extended data and fitting procedure). 
\begin{figure}[!tb]
    \centering
    \includegraphics[width=0.5\linewidth]{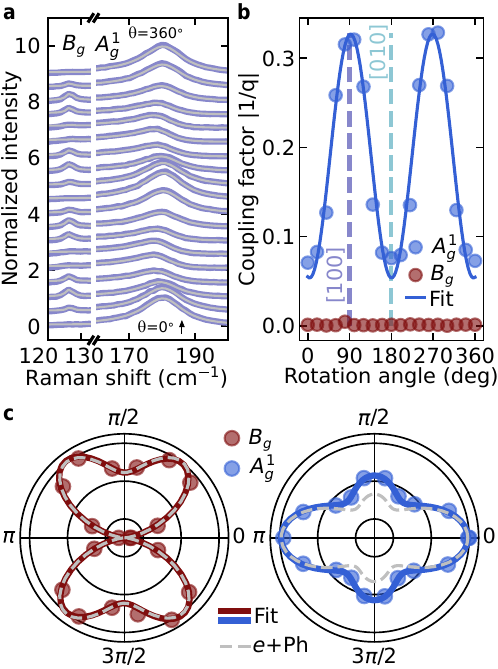}
    \caption{Characterizing the electronic continuum via angular dependence of Fano asymmetry. \textbf{a} The lineshape of the $B_g$ and $A_g^1$ peaks (purple thick line) are fitted with the sum of two Fano resonances (thin gray lines). The curves are taken for increasing in-plane polarization angle from \SI{0}{^{\circ}} to \SI{360}{^{\circ}}, normalized to the relative maximum and shifted of 0.5 for clarity. \textbf{b} Coupling factor $1/q$ for $B_g$ (red dots) and $A_g^1$ (blue dots), extracted from the fits in (a). The blue line represents a fit to the $A_g^1$. Vertical dashed lines indicate the orientation of the corresponding crystallographic axes. \textbf{c} Angular plot extracted from the ARPR spectra of Fig. \ref{fig1}d for $B_g$ (upper panel, red dots) and $A_g^1$ (lower panel, blue dots). The continuous lines represent fits to the data with Eq. S11 and S9, respectively, while the gray dashed lines show the curves obtained by removing the coherent term from the equations.}
    \label{fig2}
\end{figure}
From the fit, we extract the coupling factor $1/q$, which is proportional to the coupling strength between phonons and carriers, as a function of the in-plane polarization (Fig. \ref{fig2}b). First, we notice that the coupling factor is always negative, reflecting a phase-retarded response of the continuum relative to the driving optical field, which is compatible with the presence of free carriers in the system \cite{tan2017observation,tanwar2022fano, zhang2022fano}. The absolute value of the coupling factor $|1/q|$ for the $A_g^1$ phonon exhibits a pronounced angular modulation, increasing from $\sim 0.07$ for $\mathbf{e}_i \parallel [010]$ to $\sim0.33$ for $\mathbf{e}_i \parallel [100]$. By fitting the oscillation with the functional form $|1/q| \propto \cos(n\vartheta)$, we obtain $n = 2.04$ (blue solid line in Fig. \ref{fig2}b), corresponding to a period of $\pi$. The variation of the coupling strength provides clear evidence of the complex structure of the electronic continuum \cite{devereaux2007inelastic}. By analyzing the Raman signal away from the phonon peaks to isolate the continuum contribution, we find that inelastic light scattering from the continuum occurs exclusively for polarization along the [100] axis (see Supplementary Information S2). This behavior can be traced back to the anisotropic Fermi surface, which is continuous along [100] but exhibits gaps along [010], giving rise to a quasi-1D electron gas with no available electronic states for Raman scattering in the perpendicular direction \cite{zhao2020highly, zhang2021orbital}. In contrast, the same analysis performed on the $B_g$ peak shows that the coupling strength is much weaker than the $A_g^1$ peak (Fig. \ref{fig2}b), with an almost negligible angular modulation, indicating a markedly different interaction with the electronic continuum.

To rationalize the different angular behavior among phonon symmetries, it is necessary to account for the tensorial nature of the Raman scattering process. In particular, by integrating the ARPR curves over energy intervals around the peaks, we can reconstruct the experimental polarization dependence for each peak (Fig. \ref{fig2}c, see Supplementary Information S1). By comparing these curves with the theoretical formulas for the $A_g$ and $B_g$ modes, we find that the phononic Raman tensors alone are insufficient to accurately describe the system. To fully reproduce the angular dependence of the extracted signal, it is necessary to explicitly account for the anisotropic scattering amplitude of the electronic continuum by introducing effective Raman tensors for the two phonon modes (see Supplementary Information S2 for the derivation), given by
\begin{equation}
A^{eff}_g =
\begin{pmatrix}
ae^{i\varphi_a}+ce^{i\varphi_c} & 0 \\
0 & be^{i\varphi_b}
\end{pmatrix},
\qquad
B^{eff}_g =
\begin{pmatrix}
ce^{i\varphi_c} & fe^{i\varphi_f} \\
fe^{i\varphi_f} & 0
\end{pmatrix}
\end{equation}
where the term $ce^{i\varphi_c}$ accounts for the contribution of the anisotropic electron gas. As shown by the fits in Fig. \ref{fig2}c, the curves obtained using the effective Raman tensors accurately reproduce the experimental data, confirming the validity of this approach (fitted parameters can be found in Supplementary Information S4). In addition, the different coupling behavior observed between the $A_g$ and $B_g$ peaks is naturally captured by the effective Raman tensors. The $A_g$ modes possess a component along the [100] axis ($ae^{i\varphi_a}$), which adds coherently with the electronic contribution, giving rise to the Fano resonance. In contrast, the $B_g$ mode has no intrinsic component along the crystallographic axes, so the additional electronic term combines incoherently. This different coupling is also visible in the ARPR curves: for the $B_g$ mode, the curve is still accurately described even neglecting the mixing term $\propto\cos(\varphi_{cf})$ in Eq. S11  (gray dashed curve in the left panel of Fig. \ref{fig2}c). Indeed, the fit yields $\varphi_{cf} = \pi/2$, effectively nullifying the mixing term in the equation. For the $A_g^1$ mode, instead, removing the coupling term from Eq. S9 significantly modifies the angular dependence (gray dashed line in the right panel of Fig. \ref{fig2}c), thus highlighting the importance of the coherent interaction of these phonons with the electronic continuum. Since nonzero Raman tensor elements imply a modulation of the polarizability along the corresponding crystallographic direction, the distinct interactions observed for the two modes indicate that the electronic continuum couples coherently only to excitations featuring an oscillating dipole with a component along the [100] axis, thereby validating the description as a confined electron gas.

\begin{figure}[!t]
    \centering
    \includegraphics[width=0.9\linewidth]{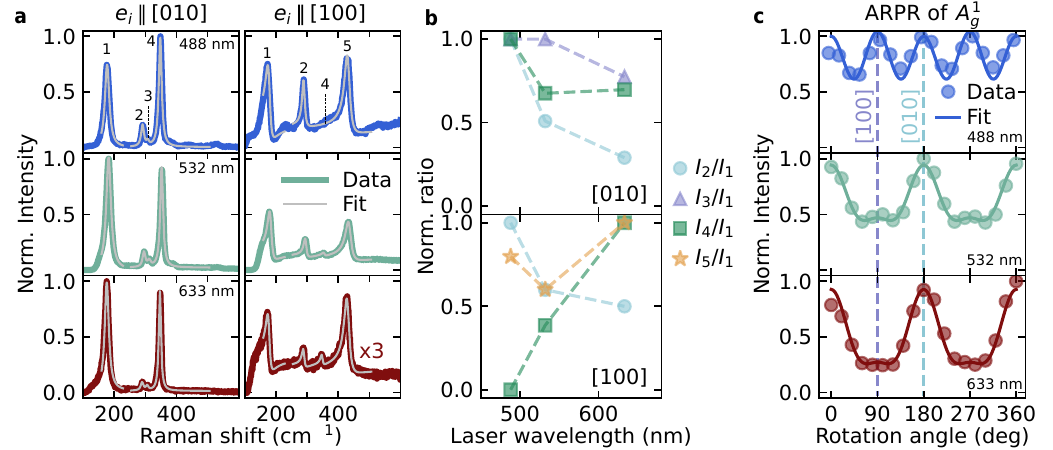}
    \caption{Evolution of Raman signal with excitation energy. \textbf{a} Raman spectra collected with $e_i\parallel[010]$ (left) and $e_i\parallel[100]$ (right) for three different excitation energies: \SI{488}{\nm}, \SI{532}{\nm} and \SI{633}{nm}, from top to bottom. Gray lines are fit to the data (see Supplementary Information S3). Data in the bottom right panel are multiplied by 3 for better visualization. \textbf{b} Ratio of intensity extracted from the fit of (a), normalized to their relative maximum. \textbf{c} Angular dispersion for peak $A_g^1$ for different excitation wavelength. Continuous line are fittings with Eq. S9.}
    \label{fig3}
\end{figure}
To gain insight on the band structure of the electronic continuum, we repeated ARPR spectroscopy with three different excitation wavelengths, spanning three different dielectric conditions: \SI{488}{\nm}, for which MoOCl\textsubscript{2} is elliptic in plane, \SI{532}{\nm}, where MoOCl\textsubscript{2} is weakly metallic along [100], and \SI{633}{\nm} where $\varepsilon_x \sim -3.2 +1i$ (see Fig. \ref{fig1}b). By normalizing the data to the relative maximum along [010] for each wavelength (Fig. \ref{fig3}a), we observe a systematic evolution of the spectra. First, the signal intensity along [100] decreases considerably relative to that along [010] as the excitation wavelength increases. This reduction reflects the decreased light penetration in the material due to the increasing metallic character of MoOCl\textsubscript{2} along the [100] axis. A second notable change is the evolution of the Fano asymmetry along the [100] axis: at lower excitation energies, the continuum appears as a smooth background, consistent with excitations of electrons within the same band (intraband electrons) \cite{burstein1971interband,riccardi2016gate}. As the excitation energy increases, the continuum develops more pronounced structure (see Supplementary Information S6 for extended data), reflecting the onset of electronic resonances associated with interband transitions or extrema in the density of states, which locally enhance coupling between vibrational modes and the electronic continuum \cite{farhat2011observation, hu2020electronic, wang2024van, gupta2025exploring}. This also explains the partial discrepancy of the Fano fit for \SI{488}{\nm} (see top right panel of Fig. \ref{fig3}a), as the Fano lineshape is strictly valid only in the case of a smooth, slowly-varying continuum.
To analyze the evolution of the scattered intensity for each peak, we normalize the intensity of peak $A_g^j$ with $j=2,\dots,5$ over the intensity of peak $A_g^1$. To isolate the relative evolution, we normalize each intensity ratio to its maximum (Fig. \ref{fig3}b). This wavelength-dependent normalization mitigates the effects of varying overall absorption while preserving potential contributions arising from differences in penetration depth (e.g., bulk-to-surface signal ratio). For polarization aligned with the [010] axis, all peaks exhibit a decreasing trend with increasing excitation energy, although the rate of decrease varies among the different modes. In contrast, peaks collected with polarization along the [100] axis exhibit distinct behaviors. In particular, $A_g^4$ disappears for higher excitation energies. A plausible explanation for this behavior is the surface sensitivity of this mode: as the excitation photon energy increases, so does the optical penetration depth, resulting in a relatively stronger bulk contribution and, consequently, an apparent reduction of the mode’s contrast and intensity. An alternative explanation may lie in the emergence of Fr{\"o}hlich interactions, namely long-range electron–phonon coupling mediated by in-plane phonons. This effect, previously investigated in other two-dimensional materials \cite{sohier2016frohlich, miller2019tuning, wu2023analyzing}, can lead to a partial renormalization of the Raman tensor components associated with modes that overlap with the electronic continuum \cite{miller2019tuning}. Such a renormalization could naturally account for the distinct intensity evolution observed for polarization along the [100] axis compared to the [010] direction, as the long-range phonon fields interact anisotropically with the electronic continuum confined along [100].
Another notable aspect of the wavelength dependence emerges from the ARPR analysis: the angular dependence of the Raman signal is strongly modulated by the excitation energy. For example, when analyzing the evolution of $A_g^1$ mode (Fig. \ref{fig3}c), the angular shape varies from nearly periodic oscillations of period $\pi/2$ for \SI{488}{\nm} to polarized scattering along [010] for \SI{633}{\nm}. Such evolution is corroborated by complementary measurements at longer excitation wavelengths reported in \cite{minnekhanov2025hyperbolic}, which confirm the same systematic trend. Similar polarization switching phenomena in Raman scattering have been reported in other two-dimensional systems \cite{pimenta2021polarized, luo2025anomalous} and have been attributed to different origins, such as resonant Raman effects \cite{mondal2025raman} or to the intrinsic anisotropy of the crystal lattice \cite{luo2025anomalous}.
The case of MoOCl\textsubscript{2} only partially falls into the latter category. Consistently with the reduction of the Raman signal along the [100] axis discussed above (Fig. \ref{fig3}a), the polarization switching effect correlates with the reduced penetration depth of the electromagnetic field along the [100] direction, leading to a progressive suppression of the $a$ tensor component as the metallic character of this axis increases. Due to the hyperbolic dielectric response of MoOCl\textsubscript{2}, this behavior also relates to the strong dielectric anisotropy of the material and cannot be fully captured within the framework of complex Raman tensors developed for anisotropic dielectric media \cite{ribeiro2015unusual, kranert2016raman}, and instead calls for an appropriate extension of the theory \cite{xie2025quantitatively}.\\
\begin{figure}[!t]
    \centering
    \includegraphics[width=0.9\linewidth]{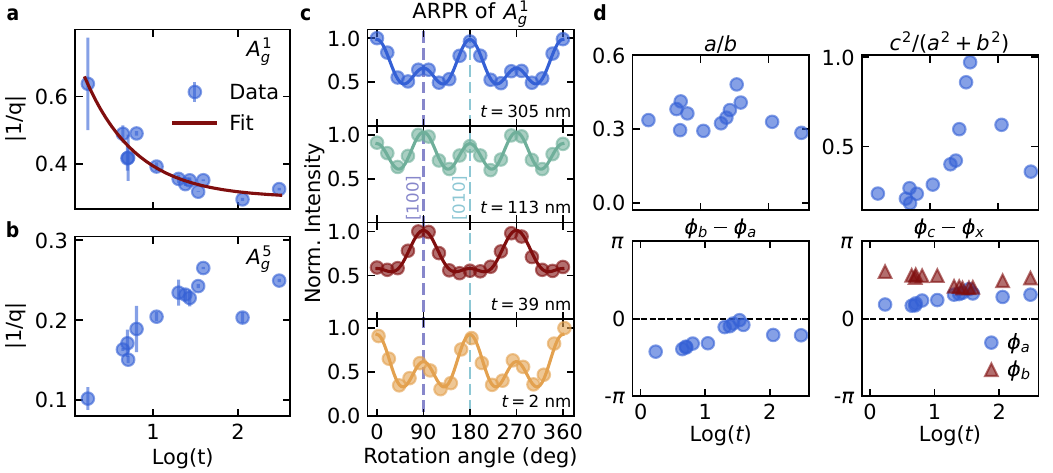}
    \caption{Coupling strength and polarization switching as a function of the thickness. \textbf{a,b} Absolute value of the coupling factor |$1/q$| as a function of the thickness $t$ of MoOCl\textsubscript{2} for $A_g^1$ (a) and $A_g^5$ (b). The reported data correspond to the average of the fitting results obtained from the two intersections of the ARPR curves with the metallic axis. The associated uncertainty is given by the semidispersion between the two measurements. The data (blue dots) in (a) are fitted with a power law $t^{-m}$ yielding $m=0.7$ (red solid line). \textbf{c} ARPR signal for $A_g^1$ excited with \SI{532}{\nm} for different thicknesses $t$ (colored dots). The continuous lines are fittings with Eqs. S10-S13. Vertical dashed lines indicate the orientation of the crystal axes. \textbf{d} Evolution of the fitted parameters.}
    \label{fig4:thickness_dep}
\end{figure}
To further characterize the electronic continuum, we can exploit the geometry of the system under examination. In general, in samples where the thickness is comparable or smaller to the field penetration depth, the amplitude of the phononic modes is naturally distributed over all layers, and therefore scales inversely with the crystal thickness \cite{froehlicher2015unified, zhang2015phonon}. Conversely, in van der Waals crystals, composed of repeated layers interconnected by weak out-of-plane van der Waals interactions, the interlayer electronic coupling can be weak \cite{barre2024engineering}, reaching the limit in which the system effectively behaves as a stack of decoupled monolayers, as observed, for example, in graphene grown on SiC \cite{sprinkle2009first}, for which the amplitude of the electronic scattering does not scale with the number of layers. In such a picture, if a phononic mode and the electronic continuum coexist in the same crystal made up of $N$ monolayers, the ratio of the mode amplitudes decreases with increasing thickness, since the phonon amplitude is distributed over all layers ($A_p\propto 1\sqrt{N}$), while the continuum coherence is localized in few layers. This difference directly affects the Fano coupling, leading to a thickness-dependent coupling factor $1/q$ that scales as $1/N^m$ with $m\in[0,1]$.\\
Thanks to this, Fano asymmetry can be used as direct probe of the effective geometry of the continuum within the crystal. By recording Raman spectra for different MoOCl\textsubscript{2} thickness, we observe that the absolute value of the maximum coupling factor $|1/q|$ reduces as a function of thickness from $\sim0.62$ for $t=$\SI{1.7}{\nm} to a value $\sim 0.33$ for $t=$\SI{305}{\nm} (blue dots in Fig. \ref{fig4:thickness_dep}a). We fitted the thickness dependence of $|1/q|$ with a power-law $\propto t^{-m}$ retrieving $m=0.7$ (see Supplementary Information S3 for the fitting procedure), which indicates that the electronic continuum in MoOCl\textsubscript{2} is predominantly confined in-plane and interacts only weakly across layers. Conversely, we observe that the coupling strength for the $A_g^5$ peak increases with the thickness (Fig. \ref{fig4:thickness_dep}b). To rationalize this seemingly counterintuitive behavior, it is instructive to examine the phonon mode displacement vectors, reported in Ref.~\cite{minnekhanov2025hyperbolic}. $A_g^1$ is a breathing mode around the Mo-O chain, which modifies the unitary cell uniformly. Thus, adding new layers displaces the mode in larger thickness with little effect on the continuum. Differently, $A_g^5$ mode is an in-plane wagging motion of the O atoms, with a dipole component along [100]. This motion strongly distorts the Mo–O chain within the unit cell, directly influencing the electronic continuum. Because the motion is mostly in-plane, adding layers does not heavily dilute its wavefunction, while the distortions accumulate, leading to an increase in the observed coupling strength with thickness. These finding, together with the pronounced in-plane confinement described above, constitutes direct experimental evidence of the quasi-1D nature of the electronic continuum in this material, arising from electrons that are predominantly displaced along the Mo–O chains and confined in the transverse in-plane and out-of-plane directions, in agreement with theoretical predictions \cite{zhao2020highly}.\\
Interestingly, thickness also modulates the ARPR signal, giving rise to polarization switching (Fig. \ref{fig4:thickness_dep}c), analogous to that observed as a function of the excitation energy (Fig. \ref{fig3}c). In particular, we observe that for \SI{532}{nm} excitation, the $A_g^1$ mode evolves from being mostly polarized along [010] for $t=$\SI{305}{\nm} (blue dots in Fig. \ref{fig4:thickness_dep}c) to scattering mainly along [100] for $t=$\SI{39}{\nm} (red dots in Fig. \ref{fig4:thickness_dep}c). The evolution of the tensor components of $A_g^1$ as a function of the thickness can be tracked by fitting the corresponding ARPR curves. For simplicity of description, we do not analyze directly the tensor elements extracted from the fit, but rather examine the evolution of two parameters: $a/b$, which gives information about the relative weight of the phonons along the two crystallographic axes, and $c^2/(a^2+b^2)$, which quantifies the relative importance of the electronic scattering and the phononic scattering (see Supplementary Information S3). We observe that the first parameter is almost constant with the thickness (upper left panel in Fig. \ref{fig4:thickness_dep}d), revealing that the relative intensity of the phonons is weakly affected by the thickness.
The second parameter shows a maximum at around \SI{40}{\nm} where the metallic response becomes comparable to the phononic contribution (upper right panel in Fig. \ref{fig4:thickness_dep}d). In addition, the $\varphi_{ba}$ phase factor evolves towards $0$ in the range from \SI{2}{nm} to \SI{39}{nm} (lower left panel in Fig. \ref{fig4:thickness_dep}d), together with an evolution of the relative phase factors between the phononic and metallic contributions (lower right panel in Fig. \ref{fig4:thickness_dep}d). This behavior is linked to the finite interaction depth of light within the material, particularly along the metallic axis. Indeed, the skin depth of MoOCl\textsubscript{2} at an excitation wavelength of \SI{532}{\nm} is $\delta=1/\Im(k)\sim$ \SI{100}{\nm}, indicating that the observed evolution can be attributed to the onset of Fabry–Pérot cavity effects or, more generally, to thickness-dependent dielectric interference phenomena. These findings highlight the central role of dielectric anisotropy in shaping the effective Raman scattering response.\\
In conclusion, our ARPR study of MoOCl\textsubscript{2} reveals a rich interplay between lattice vibrations and a strongly anisotropic electronic continuum. Pronounced Fano resonances show that $A_g$ and $B_g$ modes couple differently to this continuum. We explained this behavior by introducing an effective Raman tensor accounting for a continuum confined along [100]. The presence of such anisotropic continuum reproduces the observed angular dependence, explaining both polarization-selective Fano line shapes and mode-dependent coupling strengths. Excitation-energy–dependent measurements reveal the evolution of the continuum from smooth to structured, modulating the electron–phonon interaction and inducing non-trivial intensity variations. Moreover, we observed that the coupling strength scales with thickness, indicating that the electronic continuum interacts only weakly across layers. Taken together, our findings depict a naturally occurring quasi-1D electron gas, confined along the Mo–O chains of MoOCl\textsubscript{2}. We also observe polarization switching in the Raman response triggered by variations of the excitation energy and of the material thickness, thus revealing a complex interplay between the Raman response and the dielectric environment. These findings establish MoOCl\textsubscript{2} as the first natural hyperbolic material accessible to Raman spectroscopy, where hyperbolicity arises from a strongly anisotropic electron gas. This unique platform enables the selective probing of direction-dependent electron–phonon interactions and demonstrates significant Raman enhancement \cite{minnekhanov2025hyperbolic}, opening avenues for exploring and controlling light–matter interactions in low-dimensional hyperbolic systems.

\newpage
\printbibliography


\section*{Supplementary Information}
Additional experimental details

\section*{Acknowledgments}
 A.A. acknowledges funding from the European
Union, ERC-2025-POC 2Dchiral N.101248056. A.M. acknowledges funding from the European Union’s Horizon Europe research and innovation programme under Marie Skłodowska-Curie grant agreement no. 101146874.


\begin{refsection}

\setcounter{figure}{0}
\renewcommand{\thefigure}{S\arabic{figure}}

\setcounter{table}{0}
\renewcommand{\thetable}{S\arabic{table}}

\begin{titlepage}
\centering
{\LARGE Supplementary Information\par}
\vspace{0.5cm}
{\large Nicola Melchioni$^{1,*}$, Andrea Mancini$^1$, Antonio Ambrosio$^{1,*}$\par}
\vspace{0.5cm}
{\large \textit{$^1$Centre for Nano Science and Technology, Fondazione Istituto Italiano di Tecnologia, Via Rubattino 81, Milano 20134, Italy}\par \textit{$^*$Corresponding Authors: nicola.melchioni@iit.it, antonio.ambrosio@iit.it}}
\end{titlepage}

\section{Experimental methods}\label{SI_ExpSetup}
\subsection{Fabrication}
We employed \SI{500}{\um}-thick SiO\textsubscript{2} wafers (Wafer University) as substrates. Before starting the fabrication, the substrates were cleaned in an acetone bath for 5 minutes and rinsed with isopropyl alcohol. The flakes were obtained by mechanical exfoliation starting from a bulk crystal of MoOCl\textsubscript{2} (Nanjing MKNANO Tech. Co., Ltd.) using a commercially available processing tape (Ultron Systems, Inc). We exfoliated different samples with the scotch-tape technique, then cleaned the exfoliated substrates a second time to remove tape residues. For the thickness-dependent experiment, the thickness of the flakes was determined by employing Atomic Force Microscopy.
\subsection{Optical setup}
\begin{figure} [!h]
    \centering
    \includegraphics[width=0.7\linewidth]{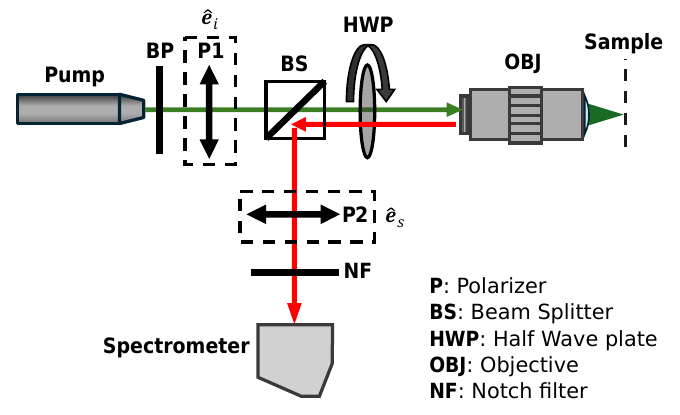}
    \caption{Schematics of the experimental setup used for this work.}
    \label{si_fig:exp_setup}
\end{figure}
The experimental setup (Fig. \ref{si_fig:exp_setup}) consists of a commercial inverted optical microscope (Nikon Eclipse Ti2) coupled to a home-built optical path.
Three different excitation energies are used: \SI{488}{\nm} (C-Wave tunable laser from Hubner), \SI{532}{\nm} (LPX-532S from Oxxius) and \SI{633}{\nm} (HeNe laser from ThorLabs). The excitation laser line is passed through a \SI{1}{nm} bandpass BP then initialized in the polarization state $\mathbf{e}_i$ by the polarizer P1. It is then passed through a 50:50 beam splitter BS and directed through a half wave plate HWP that rotates the polarization in-plane of an angle $\vartheta$. Finally, the initialized light is injected into the back aperture of a 100x Nikon objective OBJ (NA 0.9). The signal is collected in backscattering geometry with the same objective, passed through the HWP and deflected by the BS. The Raman signal projected onto the state $\mathbf{e}_s$ by the analyzer P2, isolated from the excitation signal by two notch filters and sent into an Andor Kymera 328i spectrometer. The spectrometer is equipped with a 1200 lines/mm grating, providing a spectral resolution of \SI{\sim1}{\cm^{-1}}, that projects the light onto a CMOS camera (Andor Newton) cooled down to -$70^{\circ}$C to enhance sensitivity.

\subsection{Angle-resolved measurements: simulate the rotation of the sample}
Angle-resolved polarized Raman (ARPR) measurements typically require rotating the sample under the microscope to determine the angular dependence of the Raman intensity measured at the detector. The intensity for a given phonon mode can be expressed as
\begin{equation}
    I(\vartheta) = |\mathbf{e}_s^T\cdot (M^{T}\cdot R_j\cdot M)\cdot \mathbf{e}_i|^2 
\end{equation}
where $M$ is the matrix representing an in-plane rotation of angle $\vartheta$, $R_j$ is the Raman tensor of the $j$-th mode, and $\mathbf{e}_i$ and $\mathbf{e}_s$ are the polarization vectors of the incident and scattered light, respectively.\\
In our experimental setup (Fig.~\ref{si_fig:exp_setup}), the rotation of the sample can be effectively simulated by rotating a half-wave plate (HWP) placed after the beam splitter. Specifically, the Jones matrix of a HWP with its fast axis rotated by $\vartheta/2$ relative to the laboratory $x$-axis is
\begin{equation}
J_{HWP}=
\begin{pmatrix}
\cos(\vartheta) & \sin(\vartheta) \\
\sin(\vartheta) & -\cos(\vartheta)
\end{pmatrix}\equiv M
\end{equation}
which corresponds to an in-plane rotation of linear polarization by an angle $\vartheta$. By applying the same HWP rotation to both the incident and scattered light, the relative angle between the polarization and the crystallographic axes of the sample changes, exactly reproducing the effect of physically rotating the sample under a fixed polarization.
Experimentally, the sample is placed so that the [010] axis is aligned with the horizontal polarization axis in the sample plane (0 degrees rotation of the HWP). The HWP is rotated from 0$^{\circ}$ to 180$^{\circ}$ in steps of 10$^{\circ}$ , thus equivalent to a physical rotation of the sample from 0$^{\circ}$ to 360$^{\circ}$ in 19 points. All the operations are automatized and controlled through a custom script. 
\subsection{Extraction of the experimental values for the angular dependence}
The experimental values of the angular dependence of the intensity of the peaks are extracted by integrating the spectra around each peak using the following formula:
\begin{equation}
    I_j(\vartheta) = \int_{-\gamma_j/2}^{+\gamma_j/2}{ \left[ I(\vartheta; \omega+\omega_j) - <I(0)>_{\vartheta} \right] d\omega}
\end{equation}
where $\omega_j$ and $\gamma_j$ are the central frequency and FWHM of mode $j$ (regardless if $A_g$ or $B_g$), respectively, and $<I(0)>_{\vartheta}$ is a $\vartheta$-averaged background value for $\omega<$ \SI{80}{\cm^{-1}}, where the contribution of the optical path is cut by notch filters. To extract information about the relative values of the tensor elements from the curve (see Supplementary Information \ref{si:angular_intensity_theory}), each obtained curve is normalized over its maximum.

\newpage

\section{Fano lineshape and angular dependence of the scattered intensity}\label{si:angular_intensity_theory}
\begin{figure}
    \centering
    \includegraphics[width=0.9\linewidth]{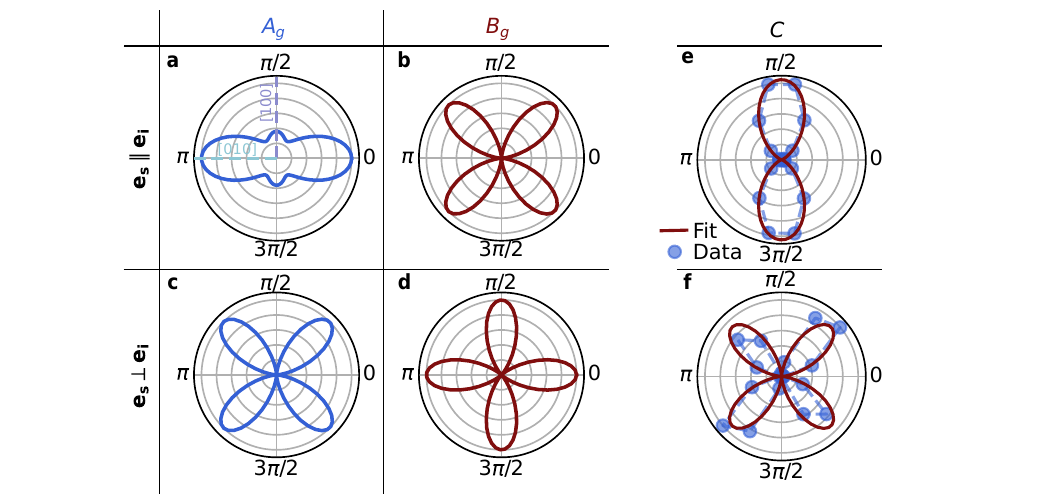}
    \caption{\textbf{a}-\textbf{d} Theoretical ARPR curves for $A_g$ (blue curves, left column) and $B_g$ (red curves, right column) modes for parallel (first row) and perpendicular (second row) configuration. Parameters for the curves are $a = 0.6$, $\varphi_a = 0$, $b = 1$, $\varphi_b = \pi/2$, $f =1 $, $\varphi_f = 0$. The dashed lines in (a) indicate the orientation of the corresponding crystal axis. \textbf{e}-\textbf{f} Angular dependence of the continuum signal away from the phononic peaks. The blue dots represent the signal extracted by integrating the signal from \SI{600}{cm^{-1}} to \SI{650}{cm^{-1}}, while the red curve is a fit computed with the tensor in Eq. \ref{si_eq:continuum_tensor}.}
    \label{si_fig:angular_theory}
\end{figure}

To reconstruct the intensity of the Raman process in MoOCl\textsubscript{2}, we start from the assumption that there is a continuum of electrons that can scatter light through Raman channels, leading to the Fano-like shape observed in the measurement. In this condition, the amplitude of the mode scattered from the surface is a superposition of the scattering from the continuum and the usual Lorentzian scattering from phonons, which can be written as
\begin{equation}
    A(\omega, \vartheta) = A_c(\omega, \vartheta)e^{i\varphi_c} + \frac{A_p(\omega, \vartheta)}{\omega -\omega_0 + i\Gamma(\vartheta)/2}
\end{equation}
 In the equation, $A_c(\omega, \vartheta)$ is a slowly-varying function of $\omega$ with real values so that the phase term of this component is only $\varphi_c$. The anisotropy of the continuum is encoded in the angular dependence. $A_p(\omega, \vartheta)$ is the complex scattered amplitude of the Raman mode.
The scattered intensity can then be written as
\begin{equation}\label{intensity_computed}
    I(\omega, \vartheta) = S(\omega, \vartheta)|A(\omega,\vartheta)|^2 + B(\omega,\vartheta)
\end{equation}
where we included a slowly-varying background $B$ and a scalar prefactor which contains information on the detector sensitivity, on the impinging laser power and other experimental factors.
As we focus on the angular dependence of each Raman peak, the explicit $\omega$ dependence can be neglected under the assumption that the measurements are performed sufficiently close to the corresponding resonance. We start by considering that in our experiment we impose a polarization for both impinging and scattered field. Thus, the amplitude of the mode observed in our experiment can be projected on the selected polarizations:
\begin{equation}\label{eq:amplitude_computed}
A (\mathbf{e_s},\mathbf{e_i}) = \mathbf{e_s}^T\cdot\left(C + R_j\right)\cdot\mathbf{e_i}
\end{equation}
where $\mathbf{e_{i,s}}$ are the polarization vectors for the impinging and scattered electric field, respectively, $C$ is the scattering tensor of the continuum, while $R_j$ is the Raman tensor of the $j$-th Raman mode.  In the experiment, we will consider only 2 possible configurations: parallel, where $\mathbf{e_s}=\mathbf{e_i}=(\cos(\vartheta),\sin(\vartheta))$, and perpendicular, where  $\mathbf{e_s}=(\sin(\vartheta),-\cos(\vartheta))$ and $\mathbf{e_i}=(\cos(\vartheta),\sin(\vartheta))$. We now need to find the correct representations for the $C$ and $R_j$ tensors.\\
The shape of the Raman tensor is assigned by the symmetries of the crystal lattice \cite{loudon1964raman}. MoOCl\textsubscript{2} belongs to the C\textit{2/m} space group, which allows for two distinct Raman-active tensor representations, $A_g$ and $B_g$ modes with tensors
\begin{equation}
A_g =
\begin{pmatrix}
ae^{i\varphi_a} & 0 \\
0 & be^{i\varphi_b}
\end{pmatrix},
\qquad
B_g =
\begin{pmatrix}
0 & fe^{i\varphi_f} \\
fe^{i\varphi_f} & 0
\end{pmatrix}
\end{equation}
where the complex phase factors are assigned to take into account the anisotropy of the crystal \cite{ribeiro2015unusual, kranert2016raman, pimenta2021polarized}. By plugging such tensors and the polarization versors into Eq. \ref{eq:amplitude_computed} with $C=0$, we can reconstruct the theoretical angular dependence of such modes, as exemplified in Fig. \ref{si_fig:angular_theory}(a-d) for selected parameter values.\\
To compute $A_c(\omega,\vartheta)$, we start from the experimental data. When analyzing the ARPR signal for parallel polarization away from the phonon peaks, we notice that the continuum of electrons scatters light only polarized along the metallic [100] axis of MoOCl\textsubscript{2} (Fig. \ref{si_fig:angular_theory}e). If we consider the [100] axis aligned with the $x$ axis of the cartesian plane, we can then write the electronic scattering tensor as
\begin{equation}\label{si_eq:continuum_tensor}
    C = \left(\begin{matrix}
        ce^{i\varphi_c} & 0\\
        0 & 0
    \end{matrix}\right)
\end{equation}
with only one component along the [100] metallic axis. By plugging this tensor into Eq. \ref{eq:amplitude_computed} with $R_j = 0$, the predicted intensities for the two conditions considered are $I_c^{\parallel}\propto \cos^4(\vartheta)$ and $I_c^{\perp}\propto \sin^2(\vartheta)\cos^2(\vartheta)$. These curves correctly reconstruct the angular dependence observed in the experiment for both polarization conditions (Fig. \ref{si_fig:angular_theory}e,f), thus confirming the accuracy of our description.

\begin{figure}[!b]
    \centering
    \includegraphics[width=0.9\linewidth]{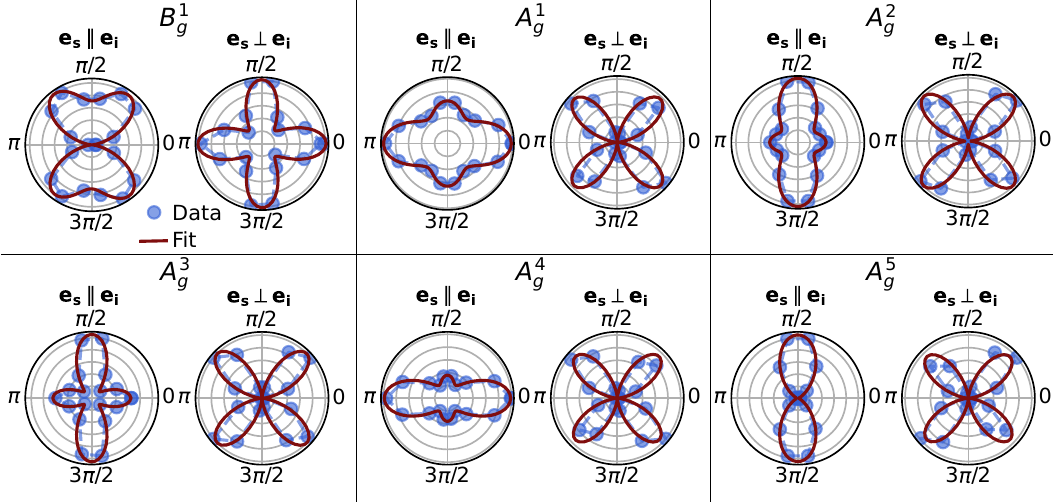}
    \caption{ARPR curves (blue dots) and fit with the corresponding equations (red solid lines) for all the peaks of MoOCl\textsubscript{2} both in parallel (left plots) and perpendicular (right plots) conditions.}
    \label{si_fig:arpr_peaks}
\end{figure}

Based on the considerations discussed above, the ARPR signal in the vicinity of the phonon peaks can be described by summing the contributions from the electronic continuum and the phonons. This leads to the definition of two effective Raman tensors corresponding to the two symmetry-allowed representations in MoOCl\textsubscript{2}, which can be written as
\begin{equation}
A^{eff}_g =
\begin{pmatrix}
ae^{i\varphi_a}+ce^{i\varphi_c} & 0 \\
0 & be^{i\varphi_b}
\end{pmatrix},
\qquad
B^{eff}_g =
\begin{pmatrix}
ce^{i\varphi_c} & fe^{i\varphi_f} \\
fe^{i\varphi_f} & 0
\end{pmatrix}
\end{equation}
By putting these Raman tensors in Eq. \ref{eq:amplitude_computed} and \ref{intensity_computed}, it is possible to compute the angular dependence of the intensity signal at the detector. We then obtain 2 different curves for each of the 2 polarization conditions, for a total of 4 intensity curves:
\begin{align}
    I_{A_g}^{\parallel}(\vartheta) &=
\cos^4(\vartheta)\left(c^2 + a^2 + 2ac\cos\varphi_{ca}\right)
+ b^2 \sin^4(\vartheta)
+ 2b \sin^2(\vartheta)\cos^2(\vartheta)
\left(c\cos\varphi_{cb} + a\cos\varphi_{ba}\right) \label{eq:I_Ag_parallel}\\ I_{A_g}^{\perp}(\vartheta) &=
\sin^2(\vartheta)\cos^2(\vartheta)
\left[
a^2 + b^2 + c^2
+ 2ac\cos\varphi_{ca}
- 2b\left(c\cos\varphi_{cb} + a\cos\varphi_{ba}\right)
\right] \label{eq:I_Ag_perp}\\
    I_{B_g}^{\parallel}(\vartheta) &=
c^2 \cos^4(\vartheta)
+ 4f^2 \sin^2(\vartheta)\cos^2(\vartheta)
+ 4cf \cos^3(\vartheta)\sin(\vartheta)\cos\varphi_{cf} \label{eq:I_Bg_parallel}\\ I_{B_g}^{\perp}(\vartheta) &=
c^2 \sin^2(\vartheta)\cos^2(\vartheta)
+ f^2 \cos^2(\vartheta)
- 2cf \cos\varphi_{cf}\,
\sin(\vartheta)\cos(\vartheta)\cos(2\vartheta) \label{eq:I_Bg_perp}
\end{align}
In all the lines, $\varphi_{ij} = \varphi_i-\varphi_j$. These equations fully describe the experimental angular dependence, as shown in Fig. \ref{si_fig:arpr_peaks}.

\newpage
\section{Notes on the fitting procedure}
\subsection{Spectral lines}
\begin{figure}[!h]
    \centering
    \includegraphics[width=0.7\linewidth]{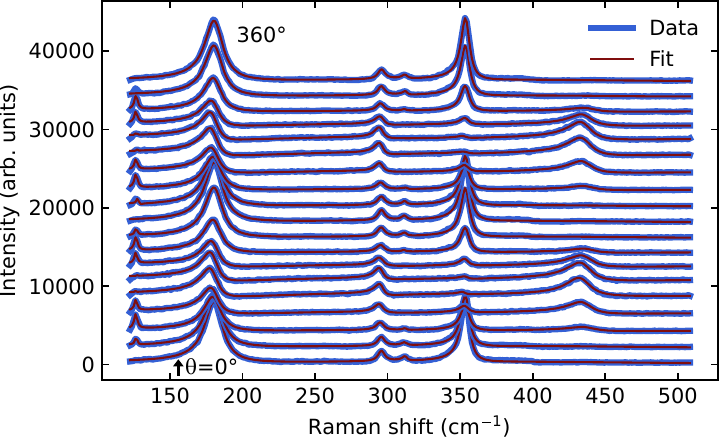}
    \caption{Raman spectra (blue curves) acquired on the \SI{305}{\nm}-thick flake at rotation angles from $\vartheta = 0^{\circ}$ (bottom) to $\vartheta = 360^{\circ}$ (top). Each spectrum was fitted using Eq. \ref{si_eq:sixfano}, and the corresponding fits are overlaid as thin red lines.}
    \label{si_fig:wl_fitting}
\end{figure}

\begin{figure}
    \centering
    \includegraphics[width=0.9\linewidth]{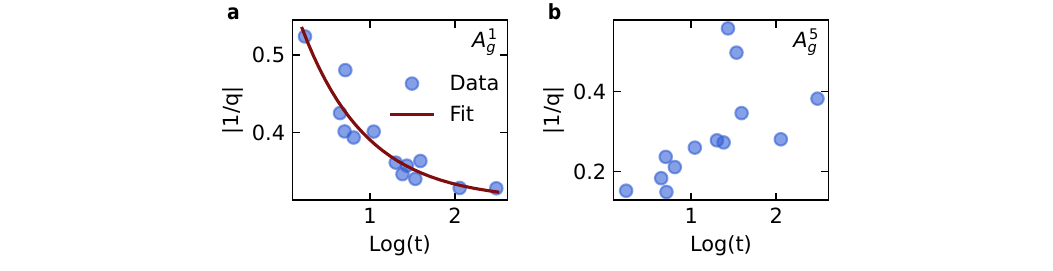}
    \caption{Coupling factor $|1/q|$ as a function of the thickness for the $A_g^1$ (a) and $A_g^2$ (b) modes, respectively. For each thickness, the point is extracted from the cosine fit of the angular dependence (blue dots). In the case of (a), the data were fitted with the power law $t^{-m}$, yielding $m=0.6$ as a result (red line).}
    \label{si_fig:method4}
\end{figure}

To extract information about the coupling strength discussed in the text, as well as peaks position, width, and intensity, the Raman spectra were fitted analytically with the sum of six Fano resonances
\begin{equation}
    I(\omega) = K+\sum_{i=1}^6a_i\frac{(\varepsilon_i+q_i)^2}{(1+\varepsilon_i^2)}, \quad \varepsilon_i = \frac{\omega-\omega_i}{\gamma_i/2}
    \label{si_eq:sixfano}
\end{equation}
where $a_i$, $q_i$, $\omega_i$, and $\gamma_i$ are the amplitude, asymmetry factor, central frequency and FWHM of the $i$-th peak, respectively, and $K$ is a global constant background. Such a model allowed the simultaneous fitting of all the peaks for each measured angle, as shown in Fig. \ref{si_fig:wl_fitting}.
\\
The coupling factors extracted from the fits are shown in Fig. 2a,b of the main text. To quantify the angular dependence of the coupling for the $A_g$ modes, we fitted the curve as
\begin{equation}
    |1/q|(\vartheta) = g_0\cos(m\vartheta+\varphi_0)+g_c
    \label{si_eq:theta_variation_cf}
\end{equation}
which results in $m\approx2$, indicating a direct correlation with the orientation of the principal axes.\\
In the case of the thickness-dependent coupling factors shown in Fig. 4a,b, we notice that the ARPR spectra intersect the [100] axis twice. We therefore extracted the coupling parameters from the fits of the spectra obtained at both crossings and report their average values, with the semidispersion taken as the associated error.\\
An alternative analysis method consists in fitting the thickness dependence of $|1/q|$ using Eq. \ref{si_eq:theta_variation_cf}, from which the maximum coupling strength is extracted as
\begin{equation}
    Max[|1/q|](t) = g_0(t)+g_c(t)
\end{equation}
This procedure ensures that the extracted parameters result from a statistical analysis of the full dataset and are not influenced by an arbitrary choice of the reference axis. The results are reported in Fig. \ref{si_fig:method4}. 
In the case of $A_g^1$, the two methods yield comparable results. The largest discrepancy is observed at$t=1.7$ nm, which is associated by a large uncertainty (see also Fig. 4a in the main text). We fitted the data with the power law $t^{-m}$, which yields $m=0.6$. In the case of $A_g^5$, both methods show similar trends. However, the $A_g^5$ peak is absent when the polarization is parallel to the [010] axis, resulting in poorly constrained fitting parameters that are not representative of the actual angular dependence and may hinder a reliable analysis. For this reason, we chose to report the first method of analysis in the main text.

\subsection{Angular-resolved polarized Raman curves}
Once the analytical expressions describing the ARPR signal of the phonon symmetries in MoOCl\textsubscript{2} are derived (see Sec. \ref{si:angular_intensity_theory}), they can be used to fit the experimental data and extract information on the Raman tensor components of each mode. To ensure the stability of the fitting procedure, it is important to minimize both the number of free parameters and the correlations among them. By inspecting Eqs.~\ref{eq:I_Ag_parallel}–\ref{eq:I_Bg_perp}, and in particular the expressions for the $A_g$ modes, two key aspects emerge. First, the tensor components $a$ and $c$ enter the same term with equal weight and are therefore intrinsically correlated. Second, the phase terms $\varphi_{ij}$ are not independent. Indeed, by defining $\Phi_1 = \varphi_b - \varphi_a$, $\Phi_2 = \varphi_c - \varphi_a$, and $\Phi_3 = \varphi_c - \varphi_b$, these quantities are linked by the relation $\Phi_1 = \Phi_2 - \Phi_3$. We exploit this constraint by fixing one of the phase differences, thereby reducing the number of free parameters in the fit. To mitigate parameter correlations, we split the fitting procedure into two steps. First, we extract the value of the $c$ component by fitting the continuum signal alone within a spectral window close to the phonon peak. We then fit the full ARPR signal of the phonon peak while fixing $c$ to the value obtained in the first step. This fitting procedure is justified under the assumption that the continuum contribution remains approximately constant over the integration window, which motivates our choice of a spectral region close to the peak for the continuum fit. From the fitting procedure, we extract the absolute values of the tensor components. However, these values depend on several factors, such as detector sensitivity, sample dielectric function, and chosen normalization. For this reason, we focus on relative quantities by defining the following parameters: $a/b$, which provides information on the relative phonon weight along the two crystallographic axes, $a/c$, which reflects the relative importance of phonon scattering compared to the contribution from the electronic continuum along [100], and $c^2/(a^2+b^2)$, which quantifies the relative impact of the electronic continuum with respect to the phononic response in the whole crystal.\\
During the fitting procedure, the chosen normalization constrains the absolute values of the tensor components to lie between 0 and 1 ($a, b, c\in[0,1]$). Moreover, the phases can be restricted to an even interval of length $2\pi$ centered at $0$, namely $\varphi_{ij}\in[-\pi, \pi]$.

\newpage
\section{Fitted parameters}
The following tables contain the parameters extracted by fitting with Eqs. \ref{eq:I_Ag_parallel}-\ref{eq:I_Bg_perp} the ARPR data collected on the \SI{305}{\nm} flake discussed in the main text and shown in Fig. \ref{si_fig:arpr_peaks}.

\begin{table}[h]
\centering
\begin{tabular}{lccc}
\hline
Mode & $f$ & $c$ & $\varphi_{cf}$ \\
\hline
$B_g$   & 0.862 & 0.831 & 1.559   \\
\hline
\end{tabular}
\caption{Fitted Raman tensor components and relative phase factors for the observed $B_g$ phonon modes.}
\label{tab:raman_tensor_params_bg}
\end{table}
\begin{table}[h]
\centering
\begin{tabular}{lccccc}
\hline
Mode & $a$ & $b$ & $c$ & $\varphi_{ca}$ & $\varphi_{cb}$ \\
\hline
$A_g^1$ & 0.375 & 0.781 & 0.804 & 1.255 & 1.197 \\
$A_g^2$ & 0.176 & 0.372 & 0.858 & 1.297 & 0.613 \\
$A_g^3$ & 0.009 & 0.427 & 1.133 & 2.258 & 3.395 \\
$A_g^4$ & 0.018 & 0.945 & 1.118 & 1.653 & 3.041 \\
$A_g^5$ & 0.190 & 0.225 & 0.818 & 1.633 & 0.299 \\
\hline
\end{tabular}
\caption{Fitted Raman tensor components and relative phase factors for the observed $A_g^j$ phonon modes.}
\label{tab:raman_tensor_params_ag}
\end{table}

\newpage
\section{Cross-polarization ARPR data}
\begin{figure}[!h]
    \centering
    \includegraphics[width=0.9\linewidth]{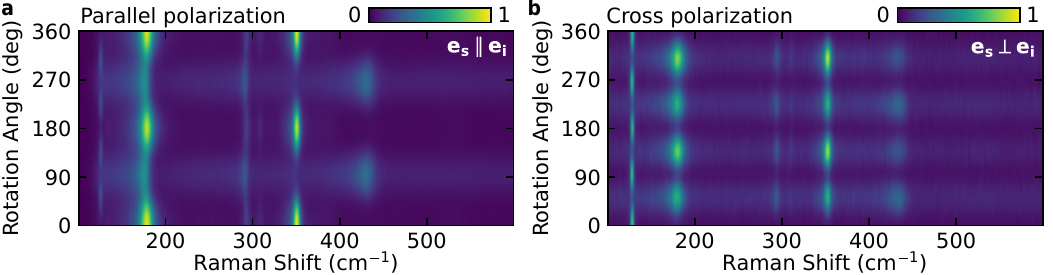}
    \caption{\textbf{a}-\textbf{b} ARPR spectra acquired on a \SI{305}{\nm} thick MoOCl\textsubscript{2} flake with excitation wavelength \SI{532}{\nm} in the parallel (a) and cross (b) polarization conditions. }
    \label{si_fig:2dparcros}
\end{figure}

In the main text, we mostly examined the $A_g^1$ and $B_g$ peaks measured in parallel polarization condition. Here, we report also the cross-polarization ARPR data. Fig. \ref{si_fig:2dparcros} reports ARPR data collected on the same \SI{305}{\nm} flake analyzed in the first part of the main text, both in the parallel polarization condition (a) and the cross polarization condition (b). In the cross polarization condition, all the peaks show a 4-lobed pattern, as shown in Sec. \ref{si:angular_intensity_theory}. However, as in parallel polarization, the different symmetry of the $B_g$ and $A_g^j$ modes is evident in cross polarization, as the maxima appear shifted of $\pi/4$. 

\newpage
\section{Wavelength dependence: extended data}
\begin{figure}[!h]
    \centering
    \includegraphics[width=0.9\linewidth]{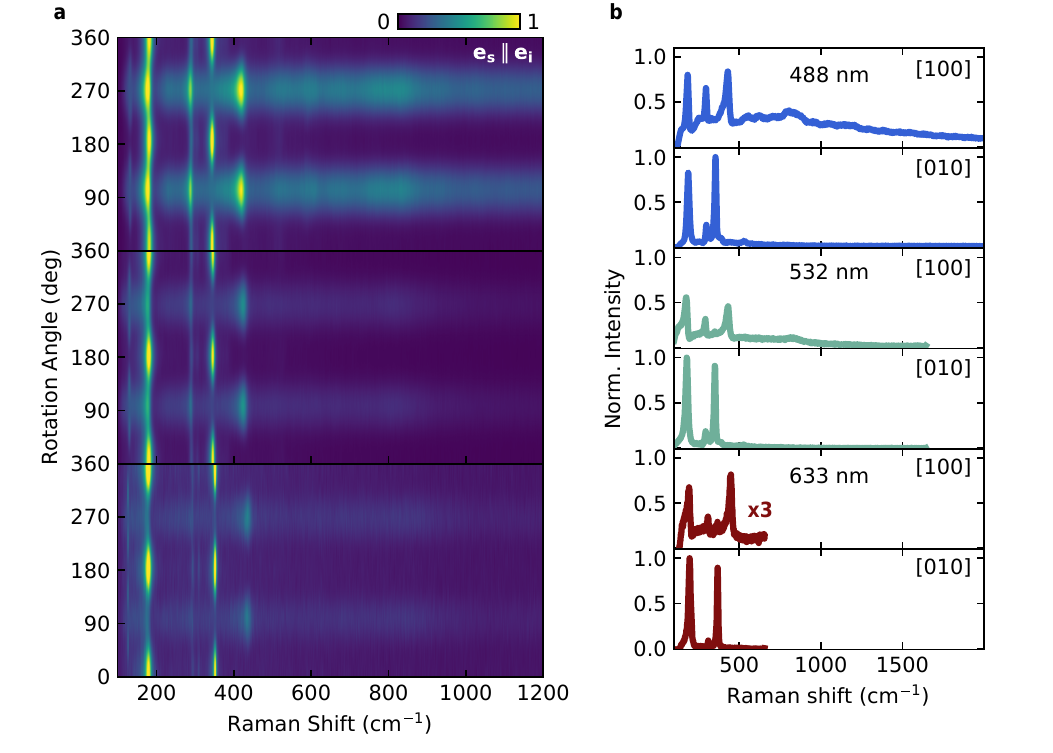}
    \caption{\textbf{a} ARPR spectra taken for parallel polarization for the three excitation wavelengths analyzed in the main text: \SI{532}{\nm} (upper panel), \SI{488}{\nm} (mid panel) and \SI{633}{\nm} (lower panel). \textbf{b} Raman spectra collected for polarization parallel to the crystallographic axes for the three excitation wavelengths extracted from (a) for the whole available spectral range.}
    \label{si_fig:wls comparison}
\end{figure}

In this section we present additional wavelength-dependent ARPR measurements. Fig. \ref{si_fig:wls comparison}a reports the ARPR spectra collected in parallel polarization for the three different excitation wavelengths used in the main text. Fig. \ref{si_fig:wls comparison}b reports the extended data used for Fig. 3a of the main text, from which the structuring of the continuum is evident.

\newpage
\section{Comparison of Fano-Fano and Lorentzian-Fano fittings of $B_g$-$A_g^1$}
\begin{figure}[!h]
    \centering
    \includegraphics[width=0.9\linewidth]{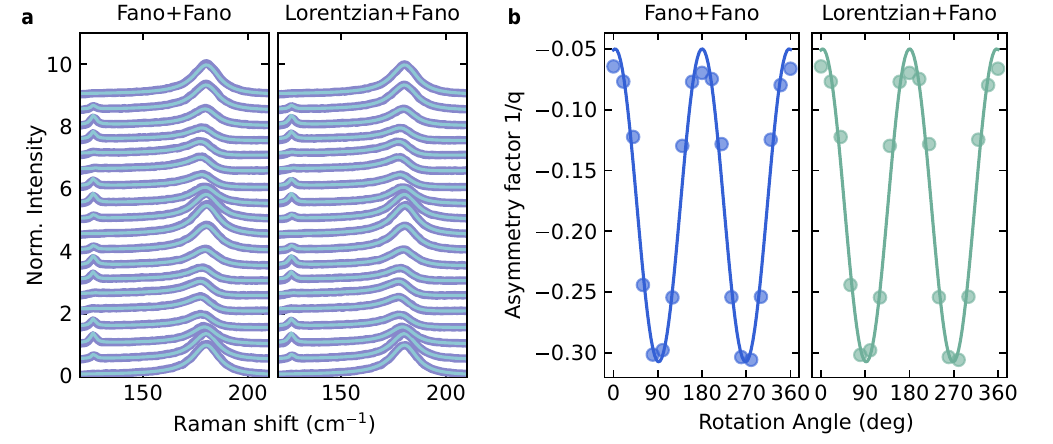}
    \caption{Comparison of two fitting methods for the $B_g$-$A_g^1$ region. \textbf{a} Raman spectra of MoOCl\textsubscript{2} as a function of the rotation angle (thick purple lines). The curves are taken for increasing in-plane polarization angle from \SI{0}{^{\circ}} to \SI{360}{^{\circ}} (from bottom to top), normalized to the relative maximum and shifted of 0.5 for clarity. The same curves are fitted with the sum of two Fano lines (left panel) and the sum of a Lorentzian and a Fano line (right panel). \textbf{b} Coupling factor $1/q$ extracted from the fits of the curves in (a) (dots). The solid lines are fits to the data with a function $\propto\cos(n\vartheta)$.}
    \label{si_fig:comparison_lorfano}
\end{figure}

In the main text (Fig. 2 and related discussion) we showed that the coupling factor of $B_g$ approaches 0, with very weak angular dependence. In this case, the lineshape is described by the limit $1/q\rightarrow0^-$ of Eq. 1 of the main text, which corresponds to a Lorentzian lineshape. It is therefore instructive to compare the result discussed in the main text with fits obtained by modeling the same spectral region as the sum of the same Fano resonance for the $A_g^1$ mode and a Lorentzian lineshape for the $B_g$ mode, described by the formula
\begin{equation}
    I(\omega) = I_0 \frac{1}{1+(\omega-\omega_0)^2/\gamma^2}
\end{equation}
As shown in Fig. \ref{si_fig:comparison_lorfano}a, the Lorentzian+Fano fit accurately reproduces the experimental data, corroborating the interpretation of the $B_g$ mode as effectively decoupled from the continuum. To verify that the description of the system is unaffected by replacing the Fano lineshape of the $B_g$ mode with a Lorentzian one, we compare the coupling factor $1/q$ extracted for the $A_g^1$ peak from the Fano+Fano fit (left panel of Fig. \ref{si_fig:comparison_lorfano}b) with that obtained from the Lorentzian+Fano fit (right panel of Fig.~\ref{si_fig:comparison_lorfano}b). No discernible differences are observed between the two cases.

\newpage

\section{Semi-analytical calculation of the ARPR curves using electric fields}
In the main text (Fig. 3b and related discussion), we showed that introducing complex tensor components and computing the ARPR curves by considering only the polarization vectors (Eqs.~\ref{eq:I_Ag_parallel}–\ref{eq:I_Bg_perp}) is insufficient to fully describe the Raman scattering in MoOCl\textsubscript{2}. Indeed, Eq. \ref{eq:amplitude_computed} already represents an approximation to the more general expression governing the Raman scattering in the material
\begin{equation}
A(\vartheta) = \mathbf{E}_s(\vartheta )^T\cdot (C+R_j)\cdot \mathbf{E_i}(\vartheta )
\end{equation}
where $\mathbf{E}_s(\vartheta)=E_s(\vartheta)\mathbf{e}_s$ and $\mathbf{E}_i=E_i(\vartheta)\mathbf{e}_i$ are the electric fields inside the material at the frequency of the scattering and excitation, respectively. When materials are isotropic, the amplitude of the fields is almost constant for every angle, and the angular dependence of the Raman response is fully captured by the relative orientation of the incident and analyzed polarization vectors with the Raman tensor, thus justifying the approximation used in Eq. \ref{eq:amplitude_computed}. In contrast, owing to the hyperbolic dielectric response of MoOCl\textsubscript{2}, the electric field inside the material is evanescent along the metallic direction and propagating along the dielectric one \cite{melchioni2025giant}. In this condition, the approximation does not hold anymore, as the strength of the Raman scattering is additionally modulated by the angle-dependent electric field distribution, and its effect must therefore be explicitly taken into account to achieve a complete description of the system.\\
\begin{figure}[!b]
    \centering
    \includegraphics[width=0.9\linewidth]{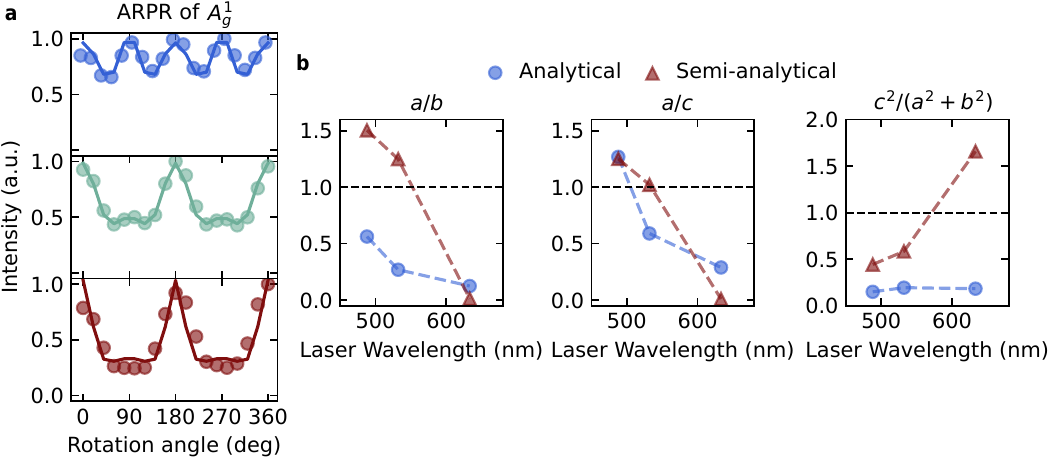}
    \caption{\textbf{a} Fitting of the ARPR experimental data (dots) with the semi-analytical formula (solid lines). \textbf{b} Comparison of the results obtained by fitting with the analytical formula (blue dots) and with the semi-analytical method with electric fields extracted from TMM (red triangles).}
    \label{si_fig:comparison_analytical_fields}
\end{figure}
Since deriving a fully analytical expression is challenging due to the complex spatial structure of the electromagnetic fields inside the material and goes beyond the scope of the present work, we developed a semi-analytical approach to improve the description of the system by employing the transfer matrix method (TMM) generalized to anisotropic layered systems \cite{passler2017generalized}. First, compute the electric field inside the MoOCl\textsubscript{2} layer at the laser and scattered frequency for different in-plane angles. We then use the obtained numerical values for $\mathbf{E}_s(\vartheta)$ and $\mathbf{E}_i(\vartheta)$ to fit the ARPR data using Eq. \ref{intensity_computed}, leaving the tensor values as free parameters. The obtained fits well reproduce the experimental data (Fig. \ref{si_fig:comparison_analytical_fields}a). When examining the fitted parameters for the $A_g^1$ mode, shown in Fig.~\ref{si_fig:comparison_analytical_fields}b, the semi-analytical model (red triangles) reproduces a decreasing trend of the $a/b$ parameter, indicating a reduced relative importance of the $a$ component with respect to the $b$ component as the metallic character of MoOCl\textsubscript{2} increases, as also observed in the analytical fit (blue dots). The same is true for the increasing contribution of the $c$ tensor component, which eventually overcomes the $a$ tensor component when the Raman scattering along the [100] axis is dominated by the electronic scattering. In contrast to the analytical fit, however, explicitly accounting for the field distributions reveal a stronger electronic response compared to the phononic one, as shown by the increased  $c^2/(a^2+b^2)$ ratio, reflecting the reduced propagation depth of the electromagnetic field inside the material and, consequently, a diminished contribution from bulk phonon scattering.\\
\begin{figure}
    \centering
    \includegraphics[width=0.9\linewidth]{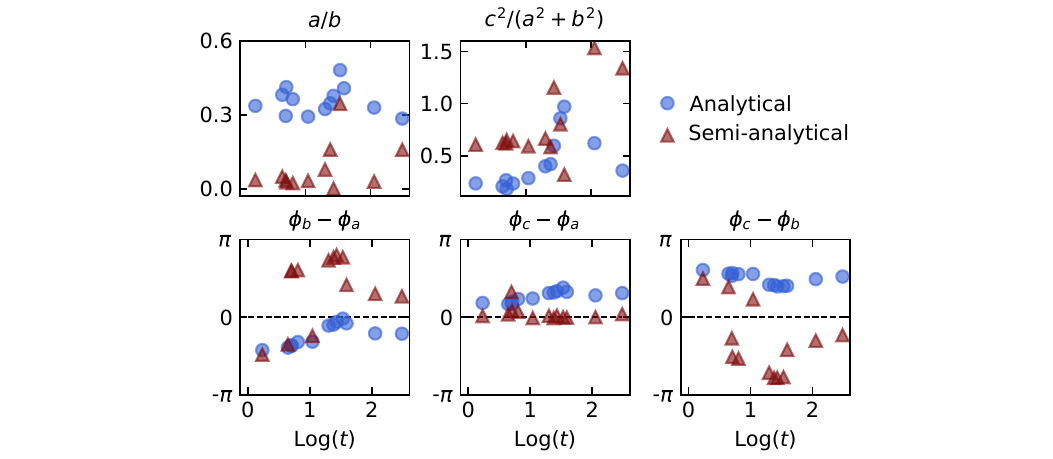}
    \caption{Comparison of the parameters extracted from the analytical fit (blue dots) and the semi-analytical fit (red triangles) for different MoOCl\textsubscript{2} thickness.}
    \label{si_fig:thickness_with_fields}
\end{figure}
When considering the thickness dependence, using the semi-analytical method for fitting the ARPR spectra yields a larger continuum contribution especially for thicknesses larger than \SI{50}{\nm}, as in the energy-dependence case. While the two models reproduce similar trends, the difference in the results underscore the complexity of the light–matter interactions in MoOCl\textsubscript{2}, arising from its exotic dielectric anisotropy. These findings call for a revision of the standard treatment of Raman scattering to explicitly incorporate the strongly angle-dependent propagation length of the electromagnetic field, which is essential for a more accurate determination of the absolute values of the Raman tensor components in this material.

\newpage
\section{Alternative normalizations for peak evolution over excitation energy}
\begin{figure}[!h]
    \centering
    \includegraphics[width=0.9\linewidth]{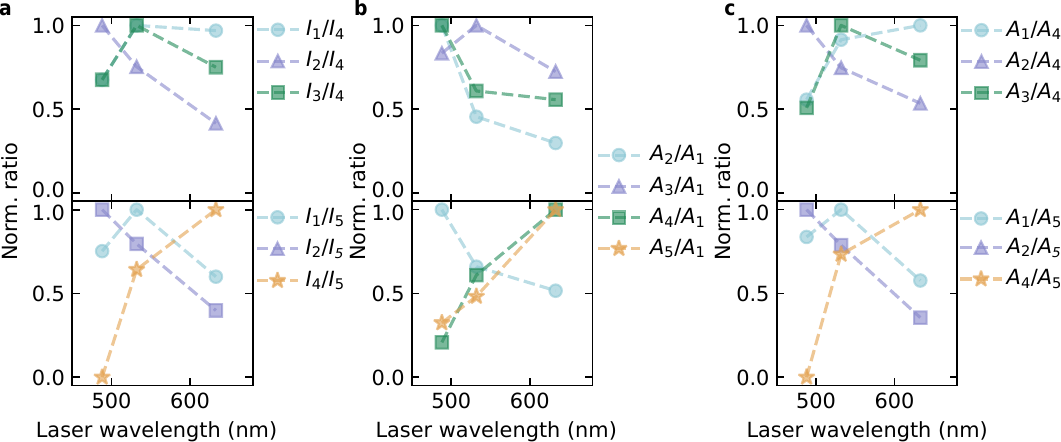}
    \caption{\textbf{a} $I^j$ with $j=1,\dots,4$ normalized over $I^4$ along the [010] direction (upper panel) and over $I^5$ along the [100] direction (lower panel). \textbf{b} Peak areas $\mathcal{A}^j$ for $j=2,\dots,5$ normalized over $\mathcal{A}^1$. \textbf{c} $\mathcal{A}^j$ with $j=1,\dots,4$ normalized over $\mathcal{A}^4$ along the [010] direction (upper panel) and over $\mathcal{A}^5$ along the [100] direction (lower panel).}
    \label{si_fig:alternative_normalizations}
\end{figure}

When studying the evolution of the peaks as a function of the excitation energy in the main text, we chose to normalize over the intensity of the $A_g^1$ mode in order to highlight the nontrivial evolution of the intensity ratios. However, such a normalization is not unique and was somewhat arbitrary. To verify that the observed trends are not artifacts, Fig. \ref{si_fig:alternative_normalizations} presents three alternative normalization schemes.
First, we normalize the intensities of the $A_g^j$ modes ($j=1,\dots,4$) over the intensity of the $A_g^4$ peak for the [010] axis (upper panel of Fig. \ref{si_fig:alternative_normalizations}a) and over the intensity of the $A_g^5$ peak for the [100] axis (lower panel of Fig. \ref{si_fig:alternative_normalizations}a). Excluding the $A_g^1$ and $A_g^5$ modes, which are used for normalization and therefore cannot be directly compared, the trends remain essentially unchanged relative to those shown in the main text.
The second alternative normalization uses the peak areas, computed as
\begin{equation}
    \mathcal{A}^j = \frac{\pi I^j\gamma^j}{2}
\end{equation}
where $I^j$ and $\gamma^j$ are the intensity and FWHM of the $j$-th peak extracted from the fit. In Fig. \ref{si_fig:alternative_normalizations}b, we show the same normalization as in the main text, but using areas instead of intensities: the areas of the $A_g^j$ modes with $j=2,\dots,5$ are normalized over the area of $A_g^1$.
Finally, Fig. \ref{si_fig:alternative_normalizations}c shows the normalization of Fig. \ref{si_fig:alternative_normalizations}a but using peak areas instead of intensities. Also in this case, the trends are essentially unchanged, confirming that the observed behavior is robust with respect to the choice of normalization.
\newpage

\printbibliography

@article{poddubny2013hyperbolic,
  title={Hyperbolic metamaterials},
  author={Poddubny, Alexander and Iorsh, Ivan and Belov, Pavel and Kivshar, Yuri},
  journal={Nature photonics},
  volume={7},
  number={12},
  pages={948--957},
  year={2013},
  publisher={Nature Publishing Group UK London}
}

@article{high2015visible,
  title={Visible-frequency hyperbolic metasurface},
  author={High, Alexander A and Devlin, Robert C and Dibos, Alan and Polking, Mark and Wild, Dominik S and Perczel, Janos and De Leon, Nathalie P and Lukin, Mikhail D and Park, Hongkun},
  journal={Nature},
  volume={522},
  number={7555},
  pages={192--196},
  year={2015},
  publisher={Nature Publishing Group UK London}
}

@article{lee2017realization,
  title={Realization of wafer-scale hyperlens device for sub-diffractional biomolecular imaging},
  author={Lee, Dasol and Kim, Yang Doo and Kim, Minkyung and So, Sunae and Choi, Hak-Jong and Mun, Jungho and Nguyen, Duc Minh and Badloe, Trevon and Ok, Jong G and Kim, Kyunghoon and others},
  journal={ACS Photonics},
  volume={5},
  number={7},
  pages={2549--2554},
  year={2017},
  publisher={ACS Publications}
}

@article{hu2020phonon,
  title={Phonon polaritons and hyperbolic response in van der Waals materials},
  author={Hu, Guangwei and Shen, Jialiang and Qiu, Cheng-Wei and Al{\`u}, Andrea and Dai, Siyuan},
  journal={Advanced Optical Materials},
  volume={8},
  number={5},
  pages={1901393},
  year={2020},
  publisher={Wiley Online Library}
}

@article{ma2018plane,
  title={In-plane anisotropic and ultra-low-loss polaritons in a natural van der Waals crystal},
  author={Ma, Weiliang and Alonso-Gonz{\'a}lez, Pablo and Li, Shaojuan and Nikitin, Alexey Y and Yuan, Jian and Mart{\'\i}n-S{\'a}nchez, Javier and Taboada-Guti{\'e}rrez, Javier and Amenabar, Iban and Li, Peining and V{\'e}lez, Sa{\"u}l and others},
  journal={Nature},
  volume={562},
  number={7728},
  pages={557--562},
  year={2018},
  publisher={Nature Publishing Group UK London}
}

@article{helmer2025computational,
  title={Computational screening of MOX 2 transition metal oxydihalides with M= V, Nb, Ta, Mo, Ru, or Os, and X= Cl, Br, or I},
  author={Helmer, Pernilla and Dahlqvist, Martin and Rosen, Johanna},
  journal={Journal of Materials Chemistry C},
  volume={13},
  number={9},
  pages={4769--4780},
  year={2025},
  publisher={Royal Society of Chemistry}
}

@book{loudon1964raman,
  title={The Raman effect in crystals},
  author={Loudon, Rodney},
  journal={Advances in Physics},
  volume={13},
  number={52},
  pages={423--482},
  year={1964},
  publisher={Taylor \& Francis}
}

@article{kranert2016raman,
  title={Raman tensor formalism for optically anisotropic crystals},
  author={Kranert, Christian and Sturm, Chris and Schmidt-Grund, R{\"u}diger and Grundmann, Marius},
  journal={Physical review letters},
  volume={116},
  number={12},
  pages={127401},
  year={2016},
  publisher={APS}
}

@article{ribeiro2015unusual,
  title={Unusual angular dependence of the Raman response in black phosphorus},
  author={Ribeiro, Henrique B and Pimenta, Marcos A and De Matos, Christiano JS and Moreira, Roberto Luiz and Rodin, Aleksandr S and Zapata, Juan D and De Souza, Eun{\'e}zio AT and Castro Neto, Antonio H},
  journal={ACS nano},
  volume={9},
  number={4},
  pages={4270--4276},
  year={2015},
  publisher={ACS Publications}
}

@article{tanwar2022fano,
  title={Fano-type wavelength-dependent asymmetric Raman line shapes from MoS2 nanoflakes},
  author={Tanwar, Manushree and Bansal, Love and Rani, Chanchal and Rani, Sonam and Kandpal, Suchita and Ghosh, Tanushree and Pathak, Devesh K and Sameera, I and Bhatia, Ravi and Kumar, Rajesh},
  journal={ACS Physical Chemistry Au},
  volume={2},
  number={5},
  pages={417--422},
  year={2022},
  publisher={ACS Publications}
}

@article{magidson2002fano,
  title={Fano-type interference in the Raman spectrum of photoexcited Si},
  author={Magidson, Valentin and Beserman, Robert},
  journal={Physical Review B},
  volume={66},
  number={19},
  pages={195206},
  year={2002},
  publisher={APS}
}

@article{pimenta2021polarized,
  title={Polarized Raman spectroscopy in low-symmetry 2D materials: angle-resolved experiments and complex number tensor elements},
  author={Pimenta, Marcos A and Resende, Geovani C and Ribeiro, Henrique B and Carvalho, Bruno R},
  journal={Physical Chemistry Chemical Physics},
  volume={23},
  number={48},
  pages={27103--27123},
  year={2021},
  publisher={Royal Society of Chemistry}
}

@article{dai2014tunable,
  title={Tunable phonon polaritons in atomically thin van der Waals crystals of boron nitride},
  author={Dai, Siyuan and Fei, Z and Ma, Q and Rodin, AS and Wagner, M and McLeod, AS and Liu, MK and Gannett, W and Regan, W and Watanabe, K and others},
  journal={Science},
  volume={343},
  number={6175},
  pages={1125--1129},
  year={2014},
  publisher={American Association for the Advancement of Science}
}

@article{margaryan2025dielectric,
  title={Dielectric permittivity tensor dynamics of in-plane hyperbolic van der Waals MoOCl2 and emergent chiral photonic applications},
  author={Margaryan, Artsruni and Sargsyan, Maksim and Hayrapetyan, Meri and Karakhanyan, David and Novoselov, Kostya S and Ghazaryan, Davit A},
  journal={arXiv preprint arXiv:2512.08383},
  year={2025}
}

@article{bechstedt1975theory,
  title={Theory of interference between electronic and phonon Raman scattering},
  author={Bechstedt, F and Peuker, K},
  journal={physica status solidi (b)},
  volume={72},
  number={2},
  pages={743--752},
  year={1975},
  publisher={Wiley Online Library}
}

@article{zhang2022fano,
  title={Fano resonance line shapes in the raman spectra of tubulin and microtubules reveal quantum effects},
  author={Zhang, Wenxu and Craddock, Travis JA and Li, Yajuan and Swartzlander, Mira and Alfano, Robert R and Shi, Lingyan},
  journal={Biophysical Reports},
  volume={2},
  number={1},
  year={2022},
  publisher={Elsevier}
}

@article{riccardi2016gate,
  title={Gate-dependent electronic Raman scattering in graphene},
  author={Riccardi, E and M{\'e}asson, M-A and Cazayous, M and Sacuto, A and Gallais, Y},
  journal={Physical Review Letters},
  volume={116},
  number={6},
  pages={066805},
  year={2016},
  publisher={APS}
}

@article{burstein1971interband,
  title={Interband electronic Raman scattering in semimetals and semiconductors},
  author={Burstein, E and Mills, DL and Wallis, RF},
  journal={Physical Review B},
  volume={4},
  number={8},
  pages={2429},
  year={1971},
  publisher={APS}
}

@article{hu2020electronic,
  title={Electronic Raman scattering in suspended semiconducting carbon nanotube},
  author={Hu, Yuecong and Chen, Shaochuang and Cong, Xin and Sun, Sida and Wu, Jiang-bin and Zhang, Daqi and Yang, Feng and Yang, Juan and Tan, Ping-Heng and Li, Yan},
  journal={The Journal of Physical Chemistry Letters},
  volume={11},
  number={24},
  pages={10497--10503},
  year={2020},
  publisher={ACS Publications}
}

@article{farhat2011observation,
  title={Observation of electronic Raman scattering in metallic carbon nanotubes},
  author={Farhat, Hootan and Berciaud, St{\'e}phane and Kalbac, Martin and Saito, Riichiro and Heinz, Tony F and Dresselhaus, Mildred S and Kong, Jing},
  journal={Physical review letters},
  volume={107},
  number={15},
  pages={157401},
  year={2011},
  publisher={APS}
}

@article{wang2024van,
  title={Van Hove Singularity-Enhanced Raman Scattering and Photocurrent Generation in Twisted Monolayer--Bilayer Graphene},
  author={Wang, Zhenlai and Zhou, Siyu and Che, Chenglong and Liu, Qiang and Zhu, Zhihong and Qin, Shiqiao and Tong, Qingjun and Zhu, Mengjian},
  journal={ACS nano},
  volume={18},
  number={36},
  pages={25183--25192},
  year={2024},
  publisher={ACS Publications}
}

@article{gupta2025exploring,
  title={Exploring electronic Raman scattering in La-doped Ce O 2: Laser energy and power-dependent Raman spectroscopy},
  author={Gupta, Minal and Rambadey, Omkar V and Aggarwal, Rahul and Sagdeo, Pankaj R},
  journal={Physical Review B},
  volume={111},
  number={23},
  pages={235208},
  year={2025},
  publisher={APS}
}

@article{wu2023analyzing,
  title={Analyzing Fundamental Properties of Two-Dimensional Materials by Raman Spectroscopy from Microscale to Nanoscale},
  author={Wu, Heng and Lin, Miao-Ling and Zhou, Yan and Zhang, Xin and Tan, Ping-Heng},
  journal={Analytical Chemistry},
  volume={95},
  number={29},
  pages={10821--10838},
  year={2023},
  publisher={ACS Publications}
}

@article{miller2019tuning,
  title={Tuning the Fr{\"o}hlich exciton-phonon scattering in monolayer MoS2},
  author={Miller, Bastian and Lindlau, Jessica and Bommert, Max and Neumann, Andre and Yamaguchi, Hisato and Holleitner, Alexander and H{\"o}gele, Alexander and Wurstbauer, Ursula},
  journal={Nature communications},
  volume={10},
  number={1},
  pages={807},
  year={2019},
  publisher={Nature Publishing Group UK London}
}

@article{mondal2025raman,
  title={Raman polarization switching in CrSBr},
  author={Mondal, Priyanka and Markina, Daria I and Hopf, Lennard and Krelle, Lukas and Shradha, Sai and Klein, Julian and Glazov, Mikhail M and Gerber, Iann and Hagmann, Kevin and Klitzing, Regine von and others},
  journal={npj 2D Materials and Applications},
  volume={9},
  number={1},
  pages={22},
  year={2025},
  publisher={Nature Publishing Group UK London}
}

@article{luo2025anomalous,
  title={Anomalous Raman Polarization Behaviors of ReS2 Thin Layers on Au Thin Film},
  author={Luo, Yu and Zhang, Jundong and Su, Weitao and Chen, Fei and Zeng, Yijie and Lu, Hong-Wei and Chen, Peiqing},
  journal={Journal of Raman Spectroscopy},
  volume={56},
  number={6},
  pages={531--537},
  year={2025},
  publisher={Wiley Online Library}
}

@article{tan2017observation,
  title={Observation of forbidden phonons, Fano resonance and dark excitons by resonance Raman scattering in few-layer WS2},
  author={Tan, Qing-Hai and Sun, Yu-Jia and Liu, Xue-Lu and Zhao, Yanyuan and Xiong, Qihua and Tan, Ping-Heng and Zhang, Jun},
  journal={2D Materials},
  volume={4},
  number={3},
  pages={031007},
  year={2017},
  publisher={IOP Publishing}
}

@article{minnekhanov2025hyperbolic,
  title={Hyperbolic-enhanced Raman scattering in van der Waals MoOCl2: from Fano resonances to picomolar detection},
  author={Minnekhanov, Anton and Tikhonowski, Gleb and Ermolaev, Georgy and Kravtsov, Konstantin V and Tselikov, Gleb and Toksumakov, Adilet and Slavich, Aleksandr and Kazantsev, Ivan and Vyshnevyy, Andrey and Kruglov, Ivan and others},
  journal={arXiv preprint arXiv:2512.17647},
  year={2025}
}

@incollection{klein2005electronic,
  title={Electronic raman scattering},
  author={Klein, Miles V},
  booktitle={Light Scattering in Solids I: Introductory Concepts},
  pages={147--204},
  year={2005},
  publisher={Springer}
}

@article{passler2017generalized,
  title={Generalized 4$\times$ 4 matrix formalism for light propagation in anisotropic stratified media: study of surface phonon polaritons in polar dielectric heterostructures},
  author={Passler, Nikolai Christian and Paarmann, Alexander},
  journal={Journal of the Optical Society of America B},
  volume={34},
  number={10},
  pages={2128--2139},
  year={2017},
  publisher={Optical Society of America}
}

@article{froehlicher2015unified,
  title={Unified description of the optical phonon modes in N-layer MoTe2},
  author={Froehlicher, Guillaume and Lorchat, Etienne and Fernique, Fran{\c{c}}ois and Joshi, Chaitanya and Molina-S{\'a}nchez, Alejandro and Wirtz, Ludger and Berciaud, St{\'e}phane},
  journal={Nano letters},
  volume={15},
  number={10},
  pages={6481--6489},
  year={2015},
  publisher={ACS Publications}
}

@article{zhang2015phonon,
  title={Phonon and Raman scattering of two-dimensional transition metal dichalcogenides from monolayer, multilayer to bulk material},
  author={Zhang, Xin and Qiao, Xiao-Fen and Shi, Wei and Wu, Jiang-Bin and Jiang, De-Sheng and Tan, Ping-Heng},
  journal={Chemical Society Reviews},
  volume={44},
  number={9},
  pages={2757--2785},
  year={2015},
  publisher={Royal Society of Chemistry}
}

@article{galiffi2024extreme,
  title={Extreme light confinement and control in low-symmetry phonon-polaritonic crystals},
  author={Galiffi, Emanuele and Carini, Giulia and Ni, Xiang and {\'A}lvarez-P{\'e}rez, Gonzalo and Yves, Simon and Renzi, Enrico Maria and Nolen, Ryan and Wasserroth, S{\"o}ren and Wolf, Martin and Alonso-Gonzalez, Pablo and others},
  journal={Nature Reviews Materials},
  volume={9},
  number={1},
  pages={9--28},
  year={2024},
  publisher={Nature Publishing Group UK London}
}

@article{sprinkle2009first,
  title={First direct observation of a nearly ideal graphene band structure},
  author={Sprinkle, Mike and Siegel, David and Hu, Yike and Hicks, J and Tejeda, Antonio and Taleb-Ibrahimi, Amina and Le Fevre, Patrick and Bertran, Fran{\c{c}}ois and Vizzini, S and Enriquez, H and others},
  journal={Physical Review Letters},
  volume={103},
  number={22},
  pages={226803},
  year={2009},
  publisher={APS}
}

@article{barre2024engineering,
  title={Engineering interlayer hybridization in van der Waals bilayers},
  author={Barr{\'e}, Elyse and Dandu, Medha and Kundu, Sudipta and Sood, Aditya and da Jornada, Felipe H and Raja, Archana},
  journal={Nature Reviews Materials},
  volume={9},
  number={7},
  pages={499--508},
  year={2024},
  publisher={Nature Publishing Group UK London}
}

@article{devereaux2007inelastic,
  title={Inelastic light scattering from correlated electrons},
  author={Devereaux, Thomas P and Hackl, Rudi},
  journal={Reviews of modern physics},
  volume={79},
  number={1},
  pages={175--233},
  year={2007},
  publisher={APS}
}

@article{xie2025quantitatively,
  title={Quantitatively Predicting Angle-Resolved Polarized Raman Intensity of Anisotropic Layered Materials},
  author={Xie, Jia-Liang and Liu, Tao and Leng, Yu-Chen and Mei, Rui and Wu, Heng and Liu, Chen-Kai and Wang, Jia-Hong and Li, Yang and Yu, Xue-Feng and Lin, Miao-Ling and others},
  journal={Advanced Materials},
  volume={37},
  number={40},
  pages={2506241},
  year={2025},
  publisher={Wiley Online Library}
}

@article{sohier2016frohlich,
  title = {Two-dimensional Fr\"ohlich interaction in transition-metal dichalcogenide monolayers: Theoretical modeling and first-principles calculations},
  author = {Sohier, Thibault and Calandra, Matteo and Mauri, Francesco},
  journal = {Phys. Rev. B},
  volume = {94},
  issue = {8},
  pages = {085415},
  numpages = {13},
  year = {2016},
  month = {Aug},
  publisher = {American Physical Society},
  doi = {10.1103/PhysRevB.94.085415},
  url = {https://link.aps.org/doi/10.1103/PhysRevB.94.085415}
}

@article{ambrosio2018selective,
  title={Selective excitation and imaging of ultraslow phonon polaritons in thin hexagonal boron nitride crystals},
  author={Ambrosio, Antonio and Tamagnone, Michele and Chaudhary, Kundan and Jauregui, Luis A and Kim, Philip and Wilson, William L and Capasso, Federico},
  journal={Light: Science \& Applications},
  volume={7},
  number={1},
  pages={27},
  year={2018},
  publisher={Nature Publishing Group UK London}
}

@article{bergeron2023probing,
  title={Probing hyperbolic and surface phonon-polaritons in 2D materials using Raman spectroscopy},
  author={Bergeron, Alaric and Gradziel, Cl{\'e}ment and Leonelli, Richard and Francoeur, S{\'e}bastien},
  journal={Nature Communications},
  volume={14},
  number={1},
  pages={4098},
  year={2023},
  publisher={Nature Publishing Group UK London}
}

@article{melchioni2025giant,
  title={Giant Optical Anisotropy in a Natural van der Waals Hyperbolic Crystal for Visible Light Low-Loss Polarization Control},
  author={Melchioni, Nicola and Mancini, Andrea and Nan, Lin and Efimova, Anastasiia and Venturi, Giacomo and Ambrosio, Antonio},
  journal={ACS nano},
  year={2025},
  publisher={ACS Publications}
}

@article{ghosh2025direct,
  title={Direct visualization of visible-light hyperbolic plasmon polaritons in real space and time},
  author={Ghosh, Atreyie and Raab, Calvin and Spellberg, Joseph L and Mohan, Aishani and King, Sarah B},
  journal={arXiv preprint arXiv:2506.13719},
  year={2025}
}

@article{wang2024planar,
  title={Planar hyperbolic polaritons in 2D van der Waals materials},
  author={Wang, Hongwei and Kumar, Anshuman and Dai, Siyuan and Lin, Xiao and Jacob, Zubin and Oh, Sang-Hyun and Menon, Vinod and Narimanov, Evgenii and Kim, Young Duck and Wang, Jian-Ping and others},
  journal={Nature communications},
  volume={15},
  number={1},
  pages={69},
  year={2024},
  publisher={Nature Publishing Group UK London}
}

@article{shekhar2014hyperbolic,
  title={Hyperbolic metamaterials: fundamentals and applications},
  author={Shekhar, Prashant and Atkinson, Jonathan and Jacob, Zubin},
  journal={Nano convergence},
  volume={1},
  number={1},
  pages={14},
  year={2014},
  publisher={Springer}
}

@article{takayama2019optics,
  title={Optics with hyperbolic materials},
  author={Takayama, Osamu and Lavrinenko, Andrei V},
  journal={Journal of the Optical Society of America B},
  volume={36},
  number={8},
  pages={F38--F48},
  year={2019},
  publisher={OSA}
}

@article{zhao2020highly,
  title={Highly anisotropic two-dimensional metal in monolayer MoOCl 2},
  author={Zhao, Jianzhou and Wu, Weikang and Zhu, Jiaojiao and Lu, Yunhao and Xiang, Bin and Yang, Shengyuan A},
  journal={Physical Review B},
  volume={102},
  number={24},
  pages={245419},
  year={2020},
  publisher={APS}
}

@article{zhang2021orbital,
  title={Orbital-selective Peierls phase in the metallic dimerized chain MoOCl 2},
  author={Zhang, Yang and Lin, Ling-Fang and Moreo, Adriana and Dagotto, Elbio},
  journal={Physical Review B},
  volume={104},
  number={6},
  pages={L060102},
  year={2021},
  publisher={APS}
}

@article{venturi2024visible,
  title={Visible-frequency hyperbolic plasmon polaritons in a natural van der Waals crystal},
  author={Venturi, Giacomo and Mancini, Andrea and Melchioni, Nicola and Chiodini, Stefano and Ambrosio, Antonio},
  journal={Nature Communications},
  volume={15},
  number={1},
  pages={9727},
  year={2024},
  publisher={Nature Publishing Group UK London}
}

@article{ruta2024good,
  title={Good plasmons in a bad metal},
  author={Ruta, Francesco L and Shao, Yinming and Acharya, Swagata and Mu, Anqi and Jo, Na Hyun and Ryu, Sae Hee and Balatsky, Daria and Su, Yifan and Pashov, Dimitar and Kim, Brian SY and others},
  journal={Science},
  volume={387},
  number={6735},
  pages={786--791},
  year={2025},
  publisher={American Association for the Advancement of Science}
}

@article{zheng2019mid,
  title={A mid-infrared biaxial hyperbolic van der Waals crystal},
  author={Zheng, Zebo and Xu, Ningsheng and Oscurato, Stefano L and Tamagnone, Michele and Sun, Fengsheng and Jiang, Yinzhu and Ke, Yanlin and Chen, Jianing and Huang, Wuchao and Wilson, William L and others},
  journal={Science advances},
  volume={5},
  number={5},
  pages={eaav8690},
  year={2019},
  publisher={American Association for the Advancement of Science}
}

\end{refsection}

\end{document}